\documentclass[twocolumn,epjc3]{svjour3}

\RequirePackage[T1]{fontenc}

\smartqed
\usepackage{amssymb}
\usepackage[utf8]{inputenc}
\RequirePackage{graphicx}
\RequirePackage{mathptmx}    
\usepackage{xcolor}
\RequirePackage[numbers,sort&compress]{natbib}
\RequirePackage[colorlinks,citecolor=blue,urlcolor=blue,linkcolor=blue]{hyperref}
\usepackage{fixltx2e}
\usepackage{arydshln}
\RequirePackage{lineno}
\usepackage{ifthen}
\usepackage{soul}
\makeatletter
\newcommand\thefontsize[1]{{#1 The current font size is: \f@size pt\par}}
\makeatother

\usepackage{lipsum}
\usepackage[printwatermark]{xwatermark}
\usepackage{pgfplots}
\usepgfplotslibrary{external} 
\newcommand{\bit}{\begin{itemize}}
\newcommand{\eit}{\end{itemize}}

\newcommand\Tstrut{\rule{0pt}{2.6ex}}         
\newcommand\Bstrut{\rule[-0.9ex]{0pt}{0pt}}

\journalname{Eur. Phys. J. C}

\begin{document}

\title{Search for steady point-like sources in the astrophysical muon neutrino flux with 8 years of IceCube data}

\onecolumn 
\author{IceCube Collaboration\thanksref{email}: M.~G.~Aartsen\thanksref{Christchurch}
\and M.~Ackermann\thanksref{Zeuthen}
\and J.~Adams\thanksref{Christchurch}
\and J.~A.~Aguilar\thanksref{BrusselsLibre}
\and M.~Ahlers\thanksref{Copenhagen}
\and M.~Ahrens\thanksref{StockholmOKC}
\and D.~Altmann\thanksref{Erlangen}
\and K.~Andeen\thanksref{Marquette}
\and T.~Anderson\thanksref{PennPhys}
\and I.~Ansseau\thanksref{BrusselsLibre}
% ...
}
\thankstext{email}{E-Mail: analysis@icecube.wisc.edu}
\authorrunning{IceCube Collaboration}
\thankstext{a}{Earthquake Research Institute, University of Tokyo, Bunkyo, Tokyo 113-0032, Japan}
\institute{
% ...
Universit\'e Libre de Bruxelles, Science Faculty CP230, B-1050 Brussels, Belgium \label{BrusselsLibre}
% ...
\and Dept.~of Physics and Astronomy, University of Canterbury, Private Bag 4800, Christchurch, New Zealand \label{Christchurch}
% ...
\and Niels Bohr Institute, University of Copenhagen, DK-2100 Copenhagen, Denmark \label{Copenhagen}
% ...
\and Erlangen Centre for Astroparticle Physics, Friedrich-Alexander-Universit\"at Erlangen-N\"urnberg, D-91058 Erlangen, Germany \label{Erlangen}
% ...
\and Department of Physics, Marquette University, Milwaukee, WI, 53201, USA \label{Marquette}
% ...
\and Oskar Klein Centre and Dept.~of Physics, Stockholm University, SE-10691 Stockholm, Sweden \label{StockholmOKC}
% ...
\and Dept.~of Physics, Pennsylvania State University, University Park, PA 16802, USA \label{PennPhys}
% ...
\and DESY, D-15738 Zeuthen, Germany \label{Zeuthen}
} 
\date{Received: date / Accepted: date}
\onecolumn
\maketitle
\twocolumn
\sloppy
\begin{abstract}
The IceCube Collaboration has observed a high-energy astrophysical neutrino flux and recently found evidence for neutrino emission from the blazar TXS 0506+056. These results open a new window into the high-energy universe. 
However, the source or sources of most of the observed flux of astrophysical neutrinos remains uncertain.

Here, a search for steady point-like neutrino sources is performed using an unbinned likelihood analysis. The method searches for a spatial accumulation of muon-neutrino events using the very high-statistics sample of about $497\,000$ neutrinos recorded by IceCube between 2009 and 2017. The median angular resolution is $\sim1^\circ$ at 1 TeV and improves to $\sim0.3^\circ$ for neutrinos with an energy of 1 PeV.
Compared to previous analyses, this search is optimized for point-like neutrino emission with the same flux-characteristics as the observed astrophysical muon-neutrino flux
and introduces an improved event-reconstruction and parametrization of the background.
The result is an improvement in sensitivity to the muon-neutrino flux compared to the previous analysis of $\sim35\%$ assuming an $E^{-2}$ spectrum.
The sensitivity on the muon-neutrino flux is at a level of $E^2 \mathrm{d} N /\mathrm{d} E = 3\cdot 10^{-13}\,\mathrm{TeV}\,\mathrm{cm}^{-2}\,\mathrm{s}^{-1}$. 

No new evidence for neutrino sources is found in a full sky scan 
and in an \emph{a priori} candidate source list that is motivated by gamma-ray observations. Furthermore, no significant excesses above background are found from populations of sub-threshold sources. The implications of the non-observation for potential source classes are discussed.
\end{abstract}

\section{Introduction}\label{sec:introduction}

Astrophysical neutrinos are thought to be produced 
by hadronic interactions of cosmic-rays with matter or radiation fields in the vicinity of their acceleration sites~\cite{Gaisser:1994yf}. 
Unlike cosmic-rays, neutrinos are not charged and are not deflected by magnetic fields and thus point back to their origin.
Moreover, since neutrinos have a relatively small interaction cross section, they can escape from the 
sources and do not suffer absorption on their way to Earth. 
Hadronic interactions of high-energy cosmic rays may also result in high-energy or very-high-energy gamma-rays. Since gamma-rays can also arise from the interaction of relativistic leptons with low-energy photons, only neutrinos are directly linked to hadronic interactions.
The most commonly assumed neutrino-flavor flux ratios in the sources result in equal or nearly equal flavor flux ratios at Earth \cite{Athar:2005wg}.
Thus about 1/3 of the astrophysical neutrinos are expected to be muon neutrinos and muon anti-neutrinos. 

In 2013, the IceCube Collaboration reported the observation of an unresolved, 
astrophysical, high-energy, all-flavor neutrino flux, consistent with isotropy, using a sample of events which begin inside the detector (‘starting events’)~\cite{hese_paper,HESEICRC2017}.  
This observation was confirmed by the measurement of an astrophysical high-energy 
muon-neutrino flux using the complementary detection channel 
of through-going muons, produced in neutrino interactions in the vicinity of the detector~\cite{CHRISPRL,diffuse_muon_paper,MUONICRC2017}. 
Track-like events from through-going muons are ideal to search for neutrino sources because of their relatively good angular resolution.
However, to date, the sources of this flux have not been identified.

In 2018, first evidence of neutrino emission from an individual source was observed for the blazar
TXS 0506+056~\cite{IceCube:2018cha,IceCube:2018dnn}.
Multi-messenger observations following up a high-energy muon neutrino event on September 22, 2017 resulted in the detection of this blazar being in flaring state. Furthermore, evidence was found for an earlier neutrino burst from the same direction between September 2014 and March 2015. However, the total neutrino flux from this source is less than 1\% of the total observed astrophysical flux. Furthermore, the stacking of the directions of known blazars has revealed no significant excess of astrophysical neutrinos at the locations of known blazars. This indicates that blazars from the 2nd Fermi-LAT AGN catalogue contribute less than about 30\% to the total observed neutrino flux assuming an unbroken power-law spectrum with spectral index of $-2.5$~\cite{aartsen_contribution_2017}. The constraint weakens to about 40\%-80\% of the total observed neutrino flux assuming a spectral index of $-2$~\cite{IceCube:2018cha}. Note that these results are model dependent and an extrapolation beyond the catalog is uncertain.
No other previous searches have revealed a significant source or source class of astrophysical neutrinos ~\cite{aartsen_search_2015,aartsen_searches_2015,adrian-martinez_first_2016,aartsen_lowering_2016,a_very_2016,psPaper,aartsen_extending_2017,aartsen_multiwavelength_2017,antares_collaboration_search_2017,aartsen_search_2017,aartsen_constraints_2017}.

Here, a search for point-like sources is presented that takes advantage of the improved event selection and reconstruction of a muon-neutrino sample developed in~\cite{diffuse_muon_paper} and the increased livetime of eight years~\cite{MUONICRC2017} between 2009 and 2017.
The best description of the sample includes a high-energy astrophysical neutrino flux given by a single power-law with a spectral index of $2.19\pm0.10$ 
and a flux normalization, at $100\,\mathrm{TeV}$, of $\Phi_{100\,\mathrm{TeV}} = 1.01^{+0.26}_{-0.23}\times 10^{-18}\,\mathrm{GeV}^{-1}\,\mathrm{cm}^{-2}\,\mathrm{s}^{-1}\,\mathrm{sr}^{-1}$, resulting in 190 to 2145 astrophysical neutrinos in the event sample. 
Compared to the previous time-integrated point source publication by IceCube~\cite{abbasi_time-integrated_2011,icecube_collaboration_search_2013,icecube_collaboration_searches_2014,aartsen_lowering_2016,psPaper}, this analysis is optimized for sources that show similar energy spectra as the measured astrophysical muon-neutrino spectrum. 
Furthermore, a high-statistics Monte Carlo parametrization of the measured data, consisting of astrophysical and atmospherical neutrinos and including systematic uncertainties, is used to model the background expectation and thus increases the sensitivity.

Within this paper, the following tests are discussed: 
1. a full sky scan for the most significant source in the Northern hemisphere, 
2. a test for a population of sub-threshold sources based on the result of the full sky scan,
3. a search based on an \emph{a priori} defined catalog of candidate objects motivated by gamma-ray observations~\cite{psPaper},
4. a test for a population of sub-threshold sources based on the result of the \emph{a priori} defined catalog search, and
5. a test of the recently observed blazar TXS 0506+056. The tests are described in Section~\ref{sec:hypothesis} and their results are given in Section~\ref{sec:results}.

\section{Data sample}\label{sec:datasample}

\begin{table*}
    \begin{tabular}{l|l|l|l|l|l|l}
        {Season} & {Start Date} & {Livetime / days} & {Events} & {Declination Range} & $ \log_{10}(E_\nu^\mathrm{astro}/\mathrm{GeV})$ {Range} & $\log_{10}(E_\nu^\mathrm{atmos}/\mathrm{GeV})$ {Range} \\ 
        \hline
        \hline
        IC59	& 2009/05/20 & 353.39 &  21411 &  $0^\circ$ – $+90^\circ$ & 3.02 -- 5.73 & 2.37 -- 4.06\\
        \hline
        IC79	& 2010/06/01 & 310.59 &  36880 & $-5^\circ$ – $+90^\circ$ & 2.96 -- 5.82 & 2.36 -- 4.04 \\
        \hline
        IC2011	& 2011/05/13  & 359.97 &  71191 & $-5^\circ$ – $+90^\circ$ & 2.89 -- 5.76 & 2.29 -- 3.98 \\
        \hline
        IC2012	& 2012/05/15 & 331.35 &        &         &           & \\ 
        IC2013	& 2013/05/02 & 360.45 &        &         &           & \\ 
        IC2014	& 2014/05/06 & 367.96 & 367590 & $-5^\circ$ – $+90^\circ$ & 2.91 -- 5.77 & 2.29 -    3.91\\ 
        IC2015	& 2015/05/18 & 356.18 &        &         &           & \\ 
        IC2016	& 2016/05/25 & 340.95 &        &         &           &    
    \end{tabular}
    \centering
    \caption{Data samples used in this analysis and some characteristics of these samples. For each sample start date, livetime, number of observed events, and energy and declination range of the event selections are given. 
    The energy range, calculated using a spectrum of atmospheric neutrinos and astrophysical neutrinos,  spans the central 90\% of the simulated events. Astrophysical neutrinos were generated using the best-fit values listed in Section~\ref{sec:introduction}. Note that livetime values slightly deviate from Ref.~\cite{diffuse_muon_paper,MUONICRC2017} as the livetime calculation has been corrected.}
    \label{tab:livetimes}
\end{table*}
 
IceCube is a cubic-kilometer neutrino detector with 5160 digital optical modules
installed on 86 cable strings in the clear ice at the  geographic South Pole between depths 
of 1450\,m and 2450\,m~\cite{icecube_paper, detector_paper}. 
The neutrino energy and directional reconstruction relies on the optical detection of Cherenkov radiation 
emitted by secondary particles produced in neutrino interactions in the surrounding ice or the nearby bedrock.
The produced Cherenkov light is detected by digital optical modules (DOMs) each consisting of a 10 inch photomultiplier tube~\cite{the_icecube_collaboration_calibration_2010}, on-board read-out electronics~\cite{DOM_MainBoard_paper} and a high-voltage board, all contained in a spherical glass pressure vessel. Light propagation within the ice can be parametrized by the scattering and absorption behavior of the antarctic ice at the South Pole~\cite{ice_paper}.
The detector construction finished in 2010.
During construction, data was taken in partial detector configurations with 59 strings (IC59) 
from May 2009 to May 2010 and with 79 strings (IC79) from May 2010 to May 2011 before IceCube became fully operational.

For events arriving from the Southern hemisphere, the trigger rate in IceCube is dominated by  atmospheric muons produced in cosmic-ray air showers.
The event selection is  restricted to the Northern hemisphere where these muons are shielded by the Earth. Additionally, events are considered  
down to $-5^\circ$ declination, where the effective overburden of ice  is sufficient to strongly attenuate the flux of atmospheric muons. 
Even after requiring reconstructed tracks from the Northern hemisphere, the event rate is dominated by mis-reconstructed atmospheric muons.
However, these mis-reconstructed events can be reduced to less than 0.3\% of the background using a careful event selection~\cite{diffuse_muon_paper,MUONICRC2017}.
As the data were taken with different partial configurations of IceCube, the details of the event selections 
are different for each season. 
At final selection level, the sample is dominated by atmospheric 
muon neutrinos from cosmic-ray air showers~\cite{diffuse_muon_paper}. These atmospheric neutrinos form an irreducible 
background to astrophysical neutrino searches and can be separated from astrophysical neutrinos on a statistical basis only.

In total, data with a livetime of 2780.85 days are analyzed containing about $497\,000$ events at the final selection level. A summary of the different sub-samples is shown in Tab.~\ref{tab:livetimes}.

The performance of the event selection can be characterized by the effective area of muon-neutrino and anti-neutrino detection, the point spread function and the central 90\% energy range of the resulting event sample. The performance is evaluated with a full detector Monte Carlo simulation~\cite{detector_paper}.

\begin{figure}
    \centering
    \input{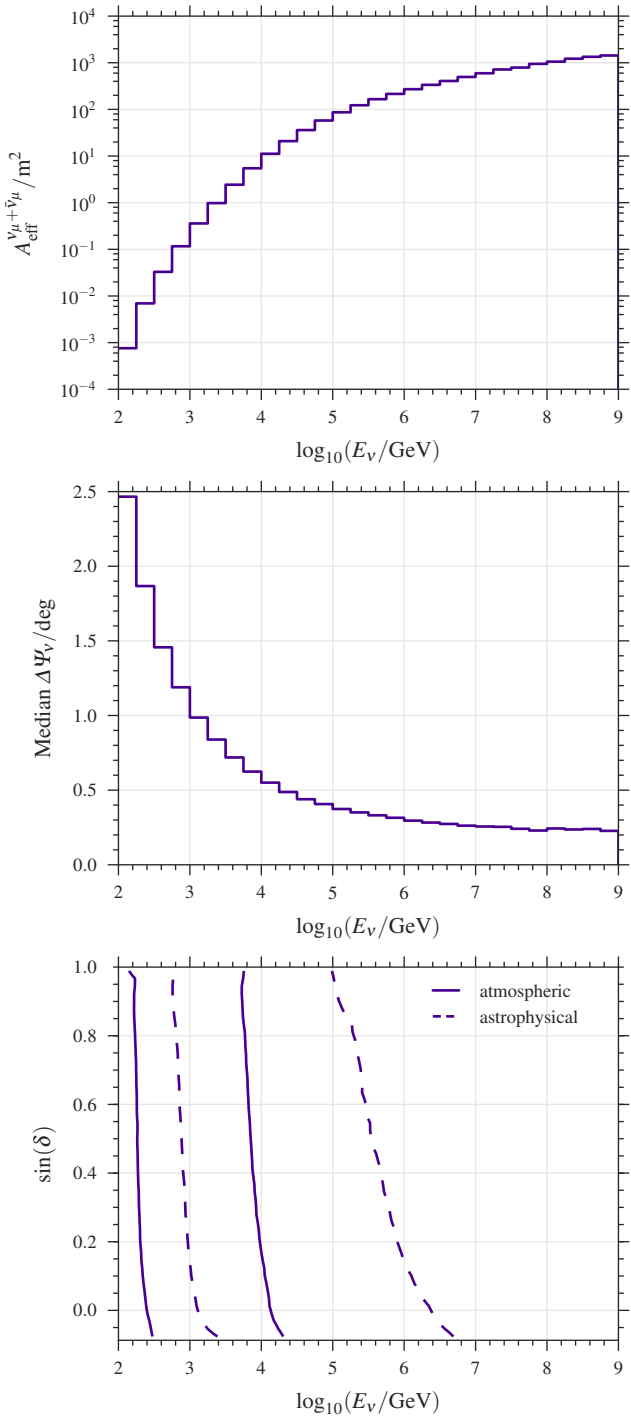}
    \caption{Top: Muon neutrino and anti-neutrino effective area averaged over the Northern hemisphere as function of $\log_{10}$ of neutrino energy. Middle: Median neutrino angular resolution as function of $\log_{10}$ of neutrino energy. Bottom: Central 90\% neutrino energy range for atmospheric (astrophysical) neutrinos as solid (dashed) line for each declination. 
    Lines show the livetime weighted averaged of all sub-samples. Plots for individual seasons can be found in the supplemental material. 
    \vspace{0.5cm}
    %\remove{For comparison, the sample labeled \emph{2012-2015} from a previous publication of time-integrated point source searches by IceCube is shown in black}~\cite{psPaper}.
    \label{fig:effective_area} \label{fig:angular_resolution} \label{fig:erange}}
\end{figure}

The effective area $A_\mathrm{eff}^{\nu+\bar{\nu}}$ quantifies the relation between neutrino and anti-neutrino fluxes $\phi_{\nu+\bar{\nu}}$ with respect to the observed rate of events $\frac{\mathrm{d}N_{\nu+\bar{\nu}}}{\mathrm{d}t}$:
\begin{equation}
    \frac{\mathrm{d}N_{\nu+\bar{\nu}}}{\mathrm{d}t} = \int \mathrm{d}\Omega \int_0^\infty \mathrm{d}E_\nu\, A_\mathrm{eff}^{\nu+\bar{\nu}}(E_\nu,\theta,\phi) \times \phi_{\nu+\bar{\nu}}(E_\nu, \theta, \phi)\,,
\end{equation}
where $\Omega$ is the solid angle, $\theta,\phi$ are the detector zenith and azimuth angle and $E_\nu$ is the neutrino energy.
The effective area for muon neutrinos and muon anti-neutrinos averaged over the Northern 
hemisphere down to -5 degree declination is shown in Fig.\ref{fig:effective_area} (top). 

At high energies, the muon direction is well correlated with the muon-neutrino direction ($< 0.1^\circ$ deviation above 10\,TeV)
and the muon is reconstructed with a median angular uncertainty $\Delta\Psi_\nu$ of about $0.6^\circ$ at $10\,\mathrm{TeV}$.
All events have been reconstructed with an improved reconstruction based on the techniques 
described in~\cite{reco_paper, Schatto14}. 
The median angular resolution is shown in Fig.\ref{fig:angular_resolution} (middle). 
The median angular resolution at a neutrino energy of $1\,\mathrm{TeV}$ is about $1^\circ$ and 
decreases for higher energies to about $0.3^\circ$ at $1\,\mathrm{PeV}$. 

The central 90\% energy range is shown in Fig.~\ref{fig:erange} (bottom) as a function of $\sin\delta$, with declination $\delta$. 
Energy ranges are calculated using the precise best-fit parametrization of the experimental sample.
The energy range stays mostly constant as function of declination but shifts to slightly higher energies near the horizon. 
The central 90\% energy range of atmospheric neutrinos is about $200\,\mathrm{GeV}$ -- $10\,\mathrm{TeV}$.

\begin{figure}
    \centering
    \input{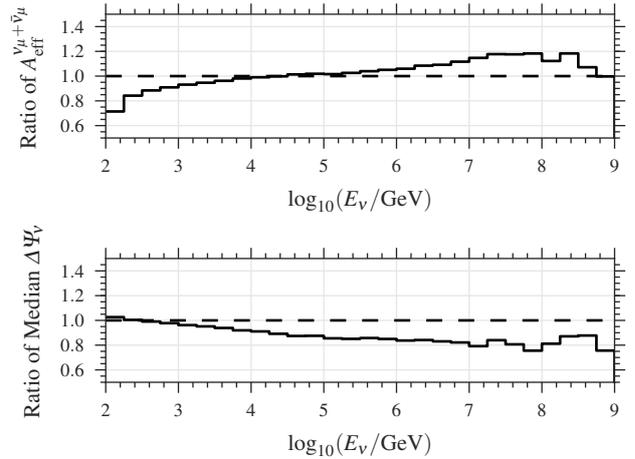}
    \caption{Ratio of effective area (top) and median angular resolution (bottom) of the sub-sample \emph{IC86 2012-2016} and the sample labeled \emph{2012-2015} from previous publication of time-integrated point source searches by IceCube is shown~\cite{psPaper}.\label{fig:ratios}}
\end{figure}

In Fig.~\ref{fig:ratios}, the ratio of effective area (top) and median angular resolution (bottom) of the sub-sample \emph{IC86 2012-2016} and the sample labeled \emph{2012-2015} from previous time-integrated point source publication by IceCube is shown~\cite{psPaper}. The differences in effective area are declination dependent. When averaged over the full Northern hemisphere, the effective area produced by this event selection is smaller than that in~\cite{psPaper} at low neutrino energies but is larger above $100\,\mathrm{TeV}$.
The median neutrino angular resolution $\Delta\Psi_\nu$ improves at $10\,\mathrm{TeV}$ by about 10\% compared to the reconstruction used in~\cite{psPaper} and improves up to 20\% at higher energies.
The event sample for the season from May 2011 to May 2012 has an overlap of about 80\% with the selection presented in Ref.~\cite{psPaper} using the same time range. 

\section{Unbinned likelihood method}\label{sec:method}

\subsection{Likelihood \& test statistics}

The data sample is tested for a spatial clustering of events with an unbinned likelihood method described in~\cite{ps_llh_paper} and used in the previous time-integrated point source publications by IceCube~\cite{abbasi_time-integrated_2011,icecube_collaboration_search_2013,icecube_collaboration_searches_2014,aartsen_lowering_2016,psPaper}. At a given position $\vec{x}_s$ in the sky, the likelihood function for a source at this position,  assuming a power law energy spectrum with spectral index $\gamma$, is given by 
\begin{equation}
	\label{eq:LLH}
    \mathcal{L} = \prod_{i}^\mathrm{events} 
                  \left[\frac{n_s}{N} S_{i}(\vec{x}_s, \gamma) + 
                  \left(1-\frac{n_s}{N}\right) B_{i} \right] 
                  \cdot P(\gamma)\,,
\end{equation}
where $i$ is an index of the observed neutrino events, $N$ is the total number of events, $n_s$ is the number of signal events and $P(\gamma)$ is a prior term.
$S_{i}$ and $B_{i}$ are the signal and background probability densities evaluated for event $i$.
The likelihood is maximized with respect to the source parameters $n_s\geq0$ and $1 \leq \gamma \leq 4$ at each tested source position in the sky given by its right ascension and declination  $\vec{x}_s=(\alpha_s, \delta_s)$.

The signal and background probability density functions (PDF) $S_i$ and $B_i$ factorize into a spatial and an energy factor
\begin{eqnarray}
    S_{i}(\vec{x}_s, \gamma) & = & S^{\mathrm{spat}}(\vec{x}_i, \sigma_i| \vec{x}_s) \cdot S^{\mathrm{ener}}(E_i | \gamma) \\
    B_{i} & = & B^{\mathrm{spat}}(\vec{x}_i) B^{\mathrm{ener}}(E_i)\,,
\end{eqnarray}
where $\vec{x}_i=(\alpha_i, \delta_i)$ is the reconstructed right ascension $\alpha_i$ and declination $\delta_i$, 
$E_i$ is the reconstructed energy~\cite{abbasi_improved_2013} and $\sigma_i$ is the event-by-event based estimated angular uncertainty of the reconstruction of event $i$~\cite{neunhoffer_estimating_2006,abbasi_time-integrated_2011}. 

 A 
likelihood ratio test is performed to compare the best-fit likelihood to the null hypothesis of no significant 
clustering $\mathcal{L}_0 = \prod_i B_{i}$. The likelihood ratio is given by
\begin{equation}
	\label{eq:TS}
    TS = 2 \cdot \log\left[\frac{\mathcal{L}(\vec{x}_s, \hat{n}_s, \hat{\gamma})}
                                {\mathcal{L}_0}\right]\,,
\end{equation}
with best-fit values $\hat{n}_s$ and $\hat{\gamma}$, which is used as a test statistic. 

The sensitivity of the analysis is defined as the median expected 90\% CL upper limit on the flux normalization in case of pure background. 
In addition, the discovery potential is defined as the signal strength that leads to a $5\,\sigma$ deviation from background in 50\% of all cases. 

In previous point source publications by IceCube~\cite{abbasi_time-integrated_2011,icecube_collaboration_search_2013,icecube_collaboration_searches_2014,aartsen_lowering_2016,psPaper}, the spatial background PDF $B^\mathrm{spat}$ and the energy background PDF $B^\mathrm{ener}$ were estimated from the data.
Given the best-fit parameters obtained from~\cite{diffuse_muon_paper} and good data / Monte Carlo agreement, it is, however, possible to get a precise parametrization of the atmospheric and diffuse astrophysical components, including systematic uncertainties. By doing this, it is possible to take advantage of the high statistics of the full detector simulation data sets which can be used to generate smooth PDFs optimized for the sample used in this work. 
Thus this parametrization of the experimental data allows us to obtain a better extrapolation to sparsely populated regions in the energy-declination plane than by using only the statistically limited experimental data. This comes with the drawback that the analysis can only be applied to the Northern hemisphere since no precise parametrization is available for the Southern hemisphere.
Generating PDFs from full detector simulations has already been done in previous publications for the energy signal PDF $S^\mathrm{ener}$, as it is not possible to estimate this PDF from data itself.
The spatial signal PDF $S^\mathrm{spat}$ is still assumed to be Gaussian with an event individual uncertainty of $\sigma_i$.

It is known from the best-fit parametrization of the sample that the data contain astrophysical events. The astrophysical component has been parametrized by an unbroken power-law with best-fit spectral index of $2.19\pm0.10$~\cite{MUONICRC2017}.
In contrast to the previous publication of time-integrated point source searches by IceCube~\cite{psPaper}, which uses a flat prior on the spectral index in the range $1 \leq \gamma \leq 4$, this analysis focuses on those sources that produce the observed spectrum of astrophysical events by adding a Gaussian prior $P(\gamma)$ on the spectral index in Eq.~\ref{eq:LLH}  with mean 2.19 and width 0.10. 
As the individual source spectra are 
not strongly constrained by the few events that contribute to a source, the prior dominates 
the fit of $\gamma$ and thus the spectral index is effectively fixed allowing only for small variations. 
Due to the prior, the likelihood has reduced effective degrees of freedom to model fluctuations. As a result, the distribution of the test statistic in the case of only background becomes steeper which results in an improvement of the discovery potential assuming an $E^{-2}$ source spectrum. 

However, due to the reduced freedom of the likelihood by the prior on the spectral index about 80\% of background trials yield $\hat{n}_s = 0$ and thus $TS=0$. This pile-up leads to an over-estimation of the median 90\% upper limit as the median is degenerate and the flux sensitivity is artificially over-estimated. 
Thus a different definition for the $TS$ is introduced for $\hat{n}_s=0$.
Allowing for negative $\hat{n}_s$ can lead to convergence problems due to the second free parameter of $\gamma$. 
Assuming $\hat{n}_s=0$ is already close to the minimum of $\log\mathcal{L}$, $\log\mathcal{L}$ can be approximated as a parabola.
The likelihood is extended in a Taylor series up to second order around $n_s=0$.
The Taylor series gives a parabola for which the value of the extremum can be calculated from the first and second order derivative of the likelihood at $n_s=0$.
This value is used as test statistic 
\begin{equation}
	\label{eq:TS2}
    TS = - 2 \cdot \frac{\left(\left.\log\mathcal{L}'\right|_{0}\right)^2}{ 2 \left.\log\mathcal{L}''\right|_{0}}\,, \qquad \hat{n}_s=0\,,
\end{equation} 
for likelihood fits that yield $\hat{n}_s=0$.
With this definition, the pile-up of $\hat{n}_s$ is spread towards negative values of $TS$ and the median of the test statistic is no longer degenerate. Using this method, the sensitivity which had been overestimated due to the pile-up at $n_s=0$ can be recovered.
 
\subsection{Pseudo-experiments}\label{sec:pseudo_experiments}

To calculate the performance of the analysis, pseudo-experiments containing
only background and pseudo-experiments with injected signal have been generated.

In this search for astrophysical point sources, atmospheric neutrinos and astrophysical neutrinos from unresolved sources make up the background. Using the precise parametrization of the reconstructed declination and energy distribution\footnote{In Ref.~\cite{MUONICRC2017}, the reconstructed zenith-energy distribution has been parametrized, although, due to IceCube's unique position at the geographic South Pole the zenith can be directly converted to declination.} from Ref.~\cite{MUONICRC2017}, pseudo-experiments are generated using full detector simulation events.
Due to IceCube's position at the South Pole and the high duty cycle of >99\%~\cite{detector_paper}, the background PDF is uniform in right ascension.

As a cross check, background samples are generated by scrambling experimental
data uniformly in right ascension. 
The declination and energy of the events are kept fixed. This results in a smaller sampled range of event energy and declination compared to the Monte Carlo-based pseudo-experiments. In the Monte Carlo-based pseudo-experiments, events are sampled from the simulated background distributions, and thus are not limited to the values of energy and declination present in the data when scrambling.
P-values for tests presented in Section~\ref{sec:results} are calculated using the Monte Carlo method and are compared to the data scrambling method for verification (values in brackets).

Signal is injected according to 
a full simulation of the detector. Events are generated at a simulated source position 
assuming a power law energy distribution.
The number of injected signal events is calculated from the assumed flux and the effective area for a small declination band around the source position.
In this analysis, the declination band was reduced compared to previous publication of time-integrated point source searches by IceCube~\cite{psPaper}, resulting in a more accurate modelling of the effective area. 
This change in signal modeling has a visible effect on the sensitivity and discovery potential, especially at the horizon and at the celestial pole. The effect can be seen in Fig.~\ref{fig:sens_bandwidth} by comparing the solid (small bandwidth) and dotted (large bandwidth) lines. 
The bandwidth is optimized by taking into account the effect of averaging over small declination bands and limited simulation statistics to calculate the effective area.
As the bandwidth cannot be made too narrow, an uncertainty of about 8\% on the flux limit calculation arises and is included in the systematic error.

\subsection{Sensitivity \& discovery potential}

\begin{figure}
   	\centering
    \input{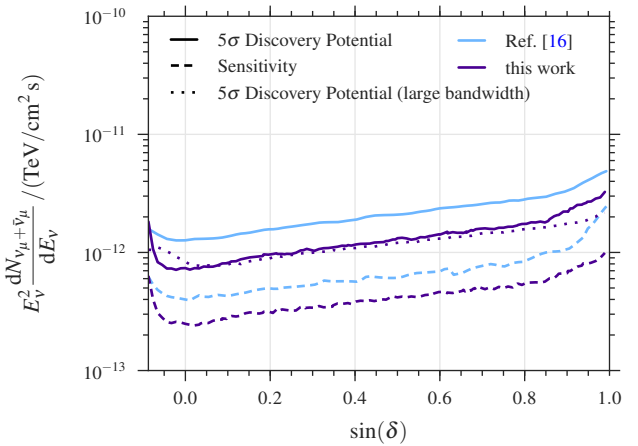}
    \caption{Sensitivity (dashed) and $5\sigma$ discovery potential (solid) of the flux normalization for an $E^{-2}$ source spectrum 
    as function of the $\sin\delta$. For comparison, the lines from~\cite{psPaper} are shown as well.
    The dotted line indicates the bandwidth effect discussed in Section~\ref{sec:pseudo_experiments}.
    }
    \label{fig:sens_bandwidth}
\end{figure}

\begin{figure}
   	\centering
    \input{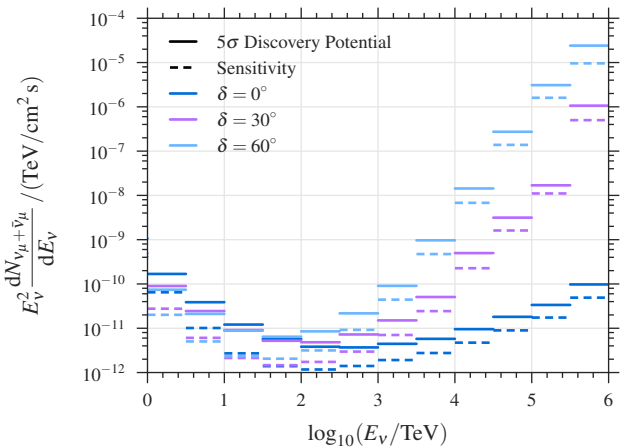}
    \caption{Differential sensitivity (dashed) and $5\sigma$ discovery potential (solid) flux for three different declinations. For high declinations and high energies, the effect of neutrino absorption within the Earth becomes visible. The flux is given as the sum of the muon neutrino and anti-neutrino flux.}
    \label{fig:differential_sensitivity_disc_pot_E2}
\end{figure}

The sensitivity and discovery potential for a single point source is calculated for an unbroken power law 
flux according to
\begin{equation}
    \frac{\mathrm{d}N_{\nu_\mu+\bar{\nu}_\mu}}{\mathrm{d}E_\nu} = \phi_{100\,\mathrm{TeV}}^{\nu_\mu+\bar{\nu}_\mu} \left(\frac{E_\nu}{100\,\mathrm{TeV}}\right)^{-\gamma}\,.
\end{equation}
In Fig.~\ref{fig:sens_bandwidth}, the sensitivity and discovery potential as function of $\sin\delta$ is shown. 
Note that Fig.~\ref{fig:sens_bandwidth} shows $E_\nu^2 \frac{\mathrm{d}N_{\nu_\mu+\bar{\nu}_\mu}}{\mathrm{d}E_\nu} = \phi_0 E_0^2$ which is constant in neutrino energy for an $E^{-2}$ flux.
The sensitivity corresponds to a 90\% CL 
averaged upper limit and the discovery potential gives the median source flux for which a $5\sigma$ discovery would be expected. The flux is given as a muon neutrino plus muon anti-neutrino flux. 
For comparison, the sensitivity and 
discovery potential from the previous publication of time-integrated point source searches by IceCube~\cite{psPaper} are shown. 
Despite only a moderate increase of livetime,
this analysis outperforms 
the analysis in~\cite{psPaper} by about 35\% for multiple reasons: 1. the use of an improved angular reconstruction, 2.
a slightly better optimized event selection near the horizon, 3. the use of background PDFs in the likelihood 
that are optimized on the parametrization from~\cite{MUONICRC2017, diffuse_muon_paper} which improves  sensitivity especially 
for higher energies, 4. the fact that due to the prior on the spectral index the number of source hypotheses 
is reduced which results in a steeper falling background $TS$ distribution, and 5. the use of negative $TS$ values which avoids overestimating the sensitivity, especially in the celestial pole region ($\sin\delta\sim1$), where the background changes rapidly in $\sin\delta$.
In Fig.~\ref{fig:differential_sensitivity_disc_pot_E2}, the differential discovery potentials for three different declination bands are shown.

\subsection{Tested hypothesis}\label{sec:hypothesis}

\subsubsection{Full sky scan}

A scan of the full Northern hemisphere from $90^\circ$ down to $-3^\circ$ declination has been performed. 
The edge at $-3^\circ$ has been chosen to avoid computational problems due to fast changing PDFs at the boundary of the sample at $-5^\circ$.
The scan is performed on a grid with a resolution of about $0.1^\circ$. The grid was generated using the HEALPix pixelization scheme\footnote{Hierarchical Equal Area isoLatitude Pixelation of a sphere (HEALPix), \url{http://healpix.sourceforge.net/}}~\cite{gorski_healpix_2005}. For each grid point, the pre-trial 
p-value is calculated.
As the test statistic shows a slight declination dependence, the declination dependent $TS$ is used to calculate local p-values. $TS$ distributions have been generated for 100 declinations equally distributed in $\sin\delta$.
$10^6$ trials have been generated for each declination. Below a $TS$ value of 5, the p-value is determined directly from trials. Above $TS=5$, an exponential function is fitted to the tail of the distribution which is used to calculate p-values above $TS=5$. A Kolmogorov-Smirnov test~\cite{kolmogorov_sulla_1933,smirnov_estimation_1939} and a $\chi^2$ test are used to verify the agreement of the fitted function and the distribution. 

The most significant point on the sky produced by the scan is selected using the pre-trial p-value.
Since many points are tested in this scan, a trial correction has to be applied. 
Therefore, the procedure is repeated with background pseudo-experiments as described in Section~\ref{sec:pseudo_experiments}. By comparing the local p-values from 
the most significant points in the background sample to the experimental pre-trial p-value, the post-trial p-value is calculated.
The final p-value is calculated directly from $\sim3500$ trials\footnote{The background distribution of the local p-value $p_\mathrm{local}$ for the most significant point is described by $d\mathcal{P} = N (1-p_\mathrm{local})^{N-1} dp_\mathrm{local}$, with an effective number of trials $N$ that is fitted to $241\,000\pm 9\,000$. A rough approximation of this trial factor can be calculated by dividing the solid angle of the Northern hemisphere $\sim 2\pi$ by the squared median angular resolution. Considering that highest energy events dominate the sensitivity, we use $0.3^\circ$ for the median angular resolution. Thus we get $2\pi/(0.3^\circ)^2 \approx 229000$ effective trials, which is in the same order of magnitude as the determined value.}.

\subsubsection{Population test in the full sky scan}

\begin{figure}[htbp]
        \centering
        \input{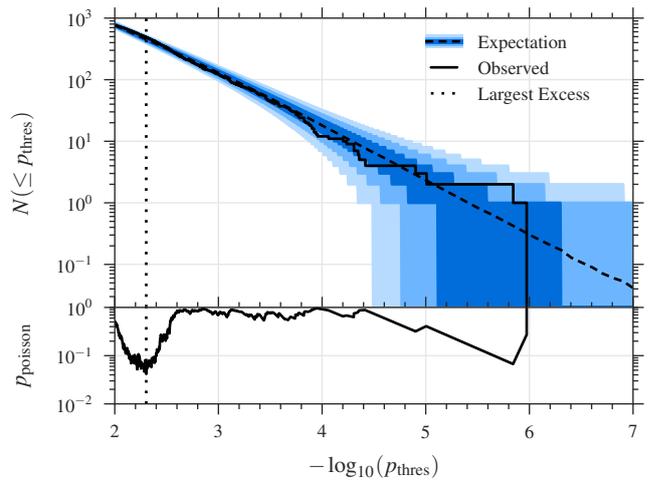}
        \caption{Upper Panel: Number of local warm spots with p-values smaller that $p_\mathrm{thres}$ as function of $p_\mathrm{thres}$. The observed number of local spots are shown as solid black line. The background expectation is shown as dashed line with $1\sigma$, $2\sigma$ and $3\sigma$ intervals corresponding to Poisson statistics. Lower Panel: Local Poisson p-value for given $p_\mathrm{thres}$. The most significant point is indicated by a dotted vertical line.}
        \label{fig:HPA_nspots_vs_pthres}
\end{figure}

Due to the large number of trials, only very strong sources would be identified in a full  sky scan, which attempts to quantify only the most significant source. However, the obtained $TS$ values can be tested also for a significant excess of events from multiple weaker sources without any bias towards source positions. This is done by counting p-values of local warm spots where the p-values are smaller than a preset threshold. An excess of counts with respect to the expectation from pure background sky maps can indicate the presence of multiple weak sources.

From the full sky scan, local spots with $p_\mathrm{local} < 10^{-2} $ and a minimal separation of $1^\circ$ are selected. The number of expected local spots $\lambda$ with a p-value smaller than $p_\mathrm{thres}$ is estimated from background pseudo-experiments and shown in Fig.~\ref{fig:HPA_nspots_vs_pthres} as dashed line. The background expectation was found to be Poisson distributed. 
The threshold value is optimized to give the most significant excess above background expectation using the Poisson probability
\begin{equation}
    p_\mathrm{poisson} = \exp(-\lambda) \sum_{m=n}^{\infty} \frac{\lambda^m}{m!}\,,
\end{equation}
to find an excess of at least $n$ spots.
Due to the optimization of the threshold in the range on $2 < -\log_{10}p_\mathrm{thres} < \infty$, the result has to be corrected for trials as well. To include this correction, the full sky scan population test is performed on background pseudo-experiments to calculate the post-trial p-value.

\subsubsection{\emph{A priori} source list}

The detectability of sources suffers from the large number of trials 
within the full sky scan and thus 
individual significant source directions
may become
insignificant after the trial correction. However, gamma-ray data can help to preselect interesting neutrino source candidates. A standard IceCube and ANTARES \emph{a priori} source 
list, containing 34 prominent candidate sources for high-energy neutrino emission on the Northern hemisphere has been tested~\cite{psPaper}, reducing the trial factor 
to about the number of sources in the catalog. The source catalog is summarized in Tab.~\ref{tab:results_source_cat}. The sources were selected mainly based on observations in gamma rays and belong to various object classes. 
The sources from this list are tested individually with the unbinned likelihood from Eq.~\ref{eq:LLH}. 
For this test, p-values are calculated from $10^6$ background trials without using any extrapolation. 
Then the most significant source is selected and a trial-correction, derived from background pseudo-experiments, is applied. 
Note that some sources such as MGRO J1908+06, SS 433, and Geminga are spatially extended with an apparent angular size of up to several degrees, which is larger than IceCube's point spread function. In such cases, the sensitivity of the analysis presented in this paper is reduced. E.g., for an extension of 1 degree, the sensitivity on the neutrino flux decreases by  $\sim20\%$~\cite{icecube_collaboration_searches_2014}.

\subsubsection{Population test in the a priori source list}

Similar to the population test in the full sky scan, an excess of several sources with small but not significant p-values in the \emph{a priori} source list can indicate a population of weak sources. 
Therefore, the $k$ most significant p-values of the source list are combined using a binomial distribution 
\begin{equation}
    P_\mathrm{binom}(k|p_k, N) = {{N}\choose{k}} p_k^k(1-p_k)^{N-k}\,,
\end{equation}
of p-values that are larger than a threshold $p_k$.
Here, $N=34$ is the total number of sources in the source list.
The most significant combination is used as a test statistic and assessed against background using pseudo-experiments.

\subsubsection{Monitored source list}

IceCube and ANTARES have tested the \emph{a priori} source list for several years with increasingly sensitive analyses~\cite{abbasi_time-integrated_2011,icecube_collaboration_search_2013,icecube_collaboration_searches_2014,psPaper}. Changing the source list posterior may lead to a bias on the result. However, not reporting on recently seen, interesting sources would also ignore progress in the field. 
A definition of an unbiased p-value is not possible as these were added later. 
Therefore, a second list with sources is tested to report on an updated source catalog. In this work, this second catalog so far comprises only TXS 0506+056, for which evidence for neutrino emission has been observed.

\subsection{Systematic uncertainties}

The p-values for the tested hypotheses are determined 
with simulated pseudo-experiments assuming only background (see also Section~\ref{sec:pseudo_experiments}).
These experiments are generated using the full detector Monte Carlo simulation, weighted to the best-fit parametrization from Ref.~\cite{MUONICRC2017}. 
This parametrization includes the optimization of nuisance parameters accounting for systematic uncertainties resulting in very good agreement between experimental data and Monte Carlo. 
Because of this procedure, the p-values 
are less affected by statistical fluctuations that would occur when estimating p-values from scrambled experimental data as well as the effect of fixed event energies during scrambling.
However, a good agreement of the parametrization with experimental data is a prerequisite of this method.
As a cross check, p-values are also calculated using scrambled experimental data. 
These p-values are given for comparison in brackets in Section~\ref{sec:results}. 
We find that the two methods show very similar results confirming the absence of systematic biases.

The calculation of the absolute neutrino flux normalization based on Monte Carlo simulations is 
affected by systematic uncertainties. These  uncertainties influence the reconstruction performance and the determination of the effective area. 
Here, the dominant uncertainties are found to be the absolute optical efficiency of the Cherenkov light production and detection in the DOMs~\cite{the_icecube_collaboration_calibration_2010}, the optical properties (absorption, scattering) of the South Pole ice~\cite{aartsen_energy_2014}, and the photo-nuclear interaction cross sections of high energy muons~\cite{bezrukov_nucleon_1981,bugaev_photonuclear_20032,bugaev_photonuclear_2003,bugaev_propagation_2004,abramowicz_parametrization_1991,abramowicz_allm_1997,koehne_proposal:_2013}. 

The systematic uncertainties on the 
sensitivity flux normalization is evaluated by propagating changed input values on the optical efficiency, ice properties and cross section values through the entire likelihood analysis 
for a signal energy spectrum of $\mathrm{d}N/\mathrm{d}E_\nu \propto E_\nu^{-2}$.
Changing the optical efficiencies by $\pm10\%$ results in a change of the flux normalization by $\pm7.5\%$. The ice properties have been varied by (+10\%, 0\%), (0\%, +10\%) and (-7.1\%, -7.1\%) in the values of absorption and scattering length. The resulting uncertainty of the flux normalization is 
$\pm5.3\%$. To study the effect of the photo-nuclear interactions of high energy muons, the models in Ref.~\cite{bezrukov_nucleon_1981,bugaev_photonuclear_20032,bugaev_photonuclear_2003,bugaev_propagation_2004,abramowicz_parametrization_1991,abramowicz_allm_1997,koehne_proposal:_2013} have been used, which give a flux normalization variation of $\pm5.1\%$. Note, that these models are outdated and represent the extreme cases from common literature. Thus, the systematic uncertainty is estimated conservatively. The systematic uncertainties are assumed to be independent and are added in quadrature, yielding a total systematic uncertainty of $\pm 10.5\%$ for the $\nu_\mu + \bar{\nu}_\mu$ flux normalization.
One should note that additionally, the modeling of point-like sources yields an uncertainty of about $\pm8\%$ as discussed in Section~\ref{sec:pseudo_experiments}.

Since the sample is assumed to be purely muon neutrino and muon anti-neutrino events, only $\nu_\mu + \bar{\nu}_\mu$ fluxes are considered. However, $\nu_\tau$ and $\bar{\nu}_\tau$ may also contribute to the observed astrophysical neutrinos in the data sample. Taking $\nu_\tau$ and $\bar{\nu}_\tau$ fluxes into account and assuming an equal flavor ratio at Earth, the sensitivity of the per-flavor flux normalization improves, depending on the declination, by 2.6\% -- 4.3\%. The expected contamination from $\nu_e$ and $\bar{\nu}_e$ is negligible.

The relative systematic uncertainty is comparable with the systematic uncertainties quoted in previous publications of time integrated point source searches by IceCube~\cite{psPaper}. In addition, the systematic effect due to the chosen finite bandwidth is included in this analysis.

\section{Results}\label{sec:results}

No significant clustering was found in any of the hypotheses tests beyond the expectation from background. Both the full-sky scan of the Northern hemisphere 
and the p-values from the source list are compatible with pure background. The p-values given in this section are calculated by pseudo-experiments based on Monte Carlo simulation weighted to the best-fit parametrization of the sample (see Section~\ref{sec:pseudo_experiments}). For  verification, p-values calculated by pseudo-experiments from scrambled experimental data are given in brackets. 

\subsection{Sky scan}

\begin{figure*}[htbp]
        \centering
        \input{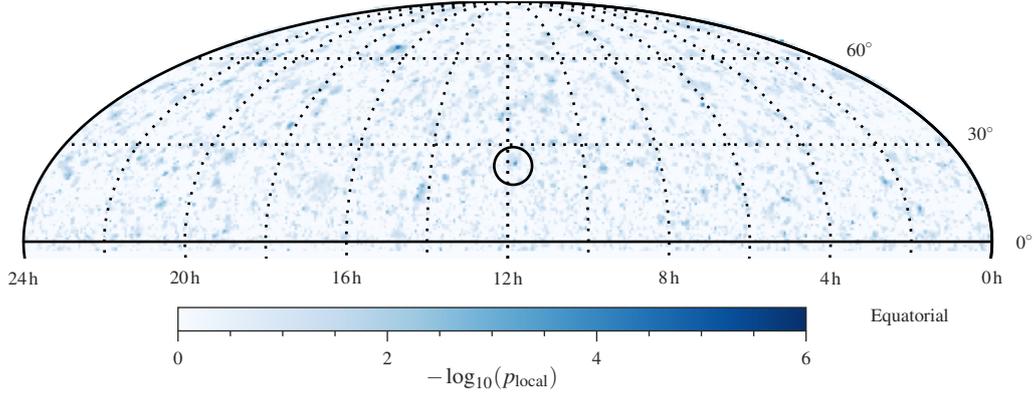}
        \caption{Sky map of the local p-values from the  sky scan in equatorial coordinates down to $-3^\circ$ declination. The local p-value is given as $-\log_{10}(p_\mathrm{local})$. The position of the most significant spot is indicated by a black circle.}
        \label{fig:sky_map_all_sky_scan_equatorial}
\end{figure*}

\begin{figure}[htbp]
        \centering
        \includegraphics{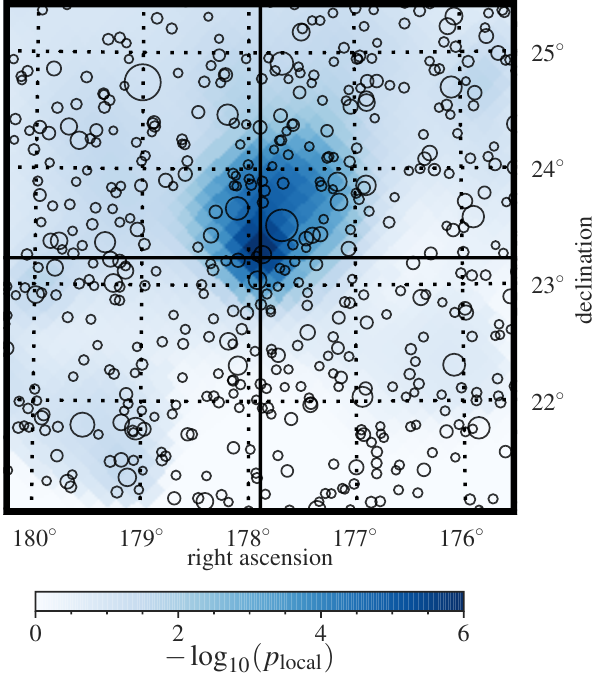}
        \caption{Local p-value landscape around the source position of the most significant spot in the  sky scan in equatorial coordinates (J2000). Neutrino event arrival directions are indicated by small circles where the area of the circles is proportional to the median $\log_{10}$ of neutrino energy assuming the diffuse best-fit spectrum. The p-value is evaluated at the point where the black lines cross.}
        \label{fig:hottest_spot_zoom}
\end{figure}

The pre-trial p-value map of the Northern hemisphere scan is shown in Fig.~\ref{fig:sky_map_all_sky_scan_equatorial}. 
The hottest spot in the scan is indicated by a black circle and is located at 
$\alpha = 177.89^\circ$ and $\delta=23.23^\circ$ (J2000) with the Galactic coordinates $b_\mathrm{gal} = 75.92^\circ$, 
$l_\mathrm{gal}=-134.33^\circ$. The best-fit signal strength is $\hat{n}_s = 21.32$ ($\Phi^{\nu_\mu+\bar{\nu}_\mu}_{100\,\mathrm{TeV}}=1.4\cdot10^{-19}\,\mathrm{GeV}^{-1}\mathrm{cm}^{-2}\mathrm{s}^{-1}$ assuming $\hat{\gamma}=2.20$) with a fitted spectral index of 
$\hat{\gamma} = 2.20$ close to the prior of $2.19$. The $TS$-value is $21.63$ which corresponds to $p_\mathrm{local}=10^{-5.97}$. 
The post-trial corrected p-value is 26.5\% (29.9\%) and is thus compatible with background. 
A zoom into the local p-value landscape around the hottest 
spot position and the observed events is shown in Fig.~\ref{fig:hottest_spot_zoom}. Events are shown as small circles where the area of the circle is proportional to the median $\log_{10}$ of neutrino energy assuming the diffuse best-fit spectrum.
The closest gamma-ray source from the Fermi 3FGL and Fermi 3FHL catalogs~\cite{3FGL, 3FHL} is 3FHL J1150.3+2418 which is about 1.1 degree away from the hottest spot. The chance probability to find a 3FGL or 3FHL source within 1.1 degree is 25\%, which is estimated from all-sky pseudo-experiments. At the source location of 3FHL J1150.3+2418, the $TS$ value is 8.02 which is inconsistent with the best-fit point at the $3.6\,\sigma$ level, if assuming Wilks theorem with one degree of freedom~\cite{wilks_large-sample_1938}.

\subsection{Population test in the  sky scan}

In Fig.~\ref{fig:HPA_nspots_vs_pthres}, the number of spots with p-values below $p_\mathrm{thres}$ are shown together with the expectation from background. 
The most significant deviation was found for $p_\mathrm{thres} = 0.5\%$ where 454.3 spots were expected and 492 were observed with a p-value of $p_\mathrm{poisson}=4.17\%$. 
Correcting the result for trials gives a p-value of 42.0\% (54.3\%) and thus the result is compatible with background.

\begin{figure}[htbp]
        \centering
        \input{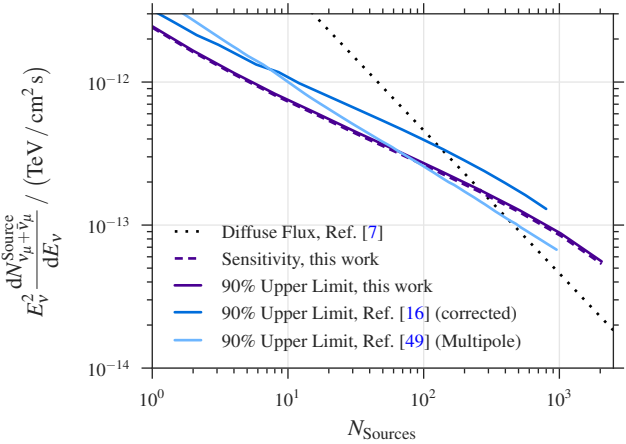}
        \caption{Single-flavor neutrino and anti-neutrino flux per source vs number of sources.
        An unbroken $E^{-2}$ power law and equal fluxes of the sources at Earth are assumed. Solid lines show 90\% CL upper limits and dashed lines indicate the sensitivity. Upper limits and sensitivity are calculated assuming that background consists of atmospheric neutrinos only and exclude an astrophysical component. Thus the limits are conservative, especially for small number of sources. For comparison, the results from~\cite{psPaper,glauchsearch2017} are given. The dotted line gives the flux per source that saturates the diffuse flux from Ref.~\cite{MUONICRC2017}.}
        \label{fig:HPA_UL_equal_strength}
\end{figure}

As no significant deviation from the background hypothesis has been observed, exclusion limits are calculated as 90\% CL upper limits  with Neyman's method~\cite{neyman_upper_limit} for the benchmark scenario of a fixed number of sources $N_\mathrm{sources}$, all producing the same flux at Earth. Upper limits are calculated assuming that background consists of atmospheric neutrinos only, excluding an astrophysical component from background pseudo-experiment generation. \label{sec:HPA_result}
Excluding the astrophysical component from background is necessary as the summed injected flux makes up a substantial part of the astrophysical flux in case of large $N_\mathrm{sources}$. However, this will over-estimate the flux sensitivity for small $N_\mathrm{sources}$. 
More realistic source scenarios are discussed in Section~\ref{sec:Kowalski}. This rather unrealistic scenario does not depend on astrophysical and cosmological assumptions about source populations and allows for a comparison between the analysis power of different analyses directly. The sensitivity and upper limits for $N_\mathrm{source}$ sources is shown in Fig.~\ref{fig:HPA_UL_equal_strength} together with the analyses from \cite{psPaper,glauchsearch2017}\footnote{The 90\% CL upper limit from Ref.~\cite{psPaper} has been recalculated to account for an incorrect treatment of signal acceptance in the original publication.}. This analysis finds the most stringent exclusion limits for small number of sources to date. The gain in sensitivity compared to Ref.~\cite{psPaper} is consistent with the gain in the sensitivity to a single point source.

\subsection{\emph{A priori} source list}\label{sec:source_list}

\begin{table*}[htbp]
\caption{Results of the \emph{a priori} defined source list search. Coordinates are given in equatorial coordinates (J2000).
  The fitted spectral index $\hat{\gamma}$ is not given as it is effectively fixed by the introduced prior. As discussed in the text, negative $TS$ values are assigned to sources with best-fit $\hat{n}_s=0$.
  Source types abbreviation: BL Lacertae object (BL Lac), Flat Spectrum Radio Quasar (FSRQ), Not Identified (NI), Pulsar Wind Nebula (PWN), Star Formation Region (SFR), Supernova Remnant (SNR), Starburst / Radio Galaxy (SRG), X-ray Binary and Micro-Quasar (XB/mqso).}
  \label{tab:results_source_cat}
  \newlength\q
  \setlength\q{\dimexpr .1\textwidth -2\tabcolsep}
  \newlength\qq
  \setlength\qq{\dimexpr .15\textwidth -2\tabcolsep}
  \newlength\qqq
  \setlength\qqq{\dimexpr .25\textwidth -2\tabcolsep}
\begin{tabular}{p{\qq}p{\q}p{\q}p{\q}p{\q}p{\q}p{\q}c}
Source & Type & $\alpha\,[\mathrm{deg}]$ & $\delta\,[\mathrm{deg}]$ & p-Value & $TS$ & $\hat{n}_s$ & $E^{2}\mathrm{d}N_{\nu_\mu+\bar{\nu}_\mu} / \mathrm{d}E\,[\mathrm{TeV}\,\mathrm{cm}^{-2}\,\mathrm{s}^{-1}]$\Tstrut\Bstrut \\
\hline \hline
4C 38.41 & FSRQ & 248.81 & 38.13 & 0.0080 & 5.0893 & 7.69 & 1.27$\cdot 10^{-12}$\\ \hline
MGRO J1908+06 & NI & 286.99 & 6.27 & 0.0088 & 4.7933 & 2.82 & 7.62$\cdot 10^{-13}$\\ \hline
Cyg A & SRG & 299.87 & 40.73 & 0.0101 & 4.7199 & 3.80 & 1.28$\cdot 10^{-12}$\\ \hline
3C454.3 & FSRQ & 343.50 & 16.15 & 0.0258 & 2.9675 & 5.03 & 8.08$\cdot 10^{-13}$\\ \hline
Cyg X-3 & XB/mqso & 308.11 & 40.96 & 0.1263 & 0.5695 & 4.33 & 8.20$\cdot 10^{-13}$\\ \hline
Cyg OB2 & SFR & 308.09 & 41.23 & 0.1706 & 0.2554 & 2.82 & 7.64$\cdot 10^{-13}$\\ \hline
LSI 303 & XB/mqso & 40.13 & 61.23 & 0.2056 & 0.1747 & 2.37 & 9.93$\cdot 10^{-13}$\\ \hline
NGC 1275 & SRG & 49.95 & 41.51 & 0.2447 & 0.0230 & 0.50 & 6.96$\cdot 10^{-13}$\\ \hline
1ES 1959+650 & BL Lac & 300.00 & 65.15 & 0.2573 & 0.0717 & 1.70 & 9.86$\cdot 10^{-13}$\\ \hline
Crab Nebula & PWN & 83.63 & 22.01 & 0.3213 & -0.0197 & 0.00 & 4.74$\cdot 10^{-13}$\\ \hline
Mrk 421 & BL Lac & 166.11 & 38.21 & 0.3460 & -0.0205 & 0.00 & 5.79$\cdot 10^{-13}$\\ \hline
Cas A & SNR & 350.85 & 58.81 & 0.3808 & -0.0169 & 0.00 & 7.01$\cdot 10^{-13}$\\ \hline
TYCHO & SNR & 6.36 & 64.18 & 0.3893 & -0.0219 & 0.00 & 7.98$\cdot 10^{-13}$\\ \hline
PKS 1502+106 & FSRQ & 226.10 & 10.52 & 0.3931 & -0.1770 & 0.00 & 3.57$\cdot 10^{-13}$\\ \hline
3C66A & BL Lac & 35.67 & 43.04 & 0.4265 & -0.1089 & 0.00 & 5.44$\cdot 10^{-13}$\\ \hline
3C 273 & FSRQ & 187.28 & 2.05 & 0.4285 & -0.3705 & 0.00 & 2.72$\cdot 10^{-13}$\\ \hline
HESS J0632+057 & XB/mqso & 98.24 & 5.81 & 0.5017 & -0.7603 & 0.00 & 2.82$\cdot 10^{-13}$\\ \hline
BL Lac & BL Lac & 330.68 & 42.28 & 0.5378 & -0.4766 & 0.00 & 4.78$\cdot 10^{-13}$\\ \hline
W Comae & BL Lac & 185.38 & 28.23 & 0.5961 & -1.0769 & 0.00 & 3.88$\cdot 10^{-13}$\\ \hline
Cyg X-1 & XB/mqso & 299.59 & 35.20 & 0.6170 & -1.0639 & 0.00 & 4.31$\cdot 10^{-13}$\\ \hline
1ES 0229+200 & BL Lac & 38.20 & 20.29 & 0.6257 & -1.6867 & 0.00 & 3.41$\cdot 10^{-13}$\\ \hline
M87 & SRG & 187.71 & 12.39 & 0.7054 & -2.9682 & 0.00 & 3.26$\cdot 10^{-13}$\\ \hline
Mrk 501 & BL Lac & 253.47 & 39.76 & 0.7214 & -1.9858 & 0.00 & 4.58$\cdot 10^{-13}$\\ \hline
PKS 0235+164 & BL Lac & 39.66 & 16.62 & 0.7494 & -3.5951 & 0.00 & 3.33$\cdot 10^{-13}$\\ \hline
H 1426+428 & BL Lac & 217.14 & 42.67 & 0.7587 & -2.5100 & 0.00 & 4.86$\cdot 10^{-13}$\\ \hline
PKS 0528+134 & FSRQ & 82.73 & 13.53 & 0.7788 & -4.4554 & 0.00 & 3.18$\cdot 10^{-13}$\\ \hline
S5 0716+71 & BL Lac & 110.47 & 71.34 & 0.7802 & -2.0711 & 0.00 & 8.02$\cdot 10^{-13}$\\ \hline
Geminga & PWN & 98.48 & 17.77 & 0.7950 & -4.7785 & 0.00 & 3.41$\cdot 10^{-13}$\\ \hline
SS433 & XB/mqso & 287.96 & 4.98 & 0.8455 & -8.0055 & 0.00 & 2.71$\cdot 10^{-13}$\\ \hline
M82 & SRG & 148.97 & 69.68 & 0.8456 & -3.5574 & 0.00 & 8.04$\cdot 10^{-13}$\\ \hline
3C 123.0 & SRG & 69.27 & 29.67 & 0.9056 & -8.2916 & 0.00 & 4.11$\cdot 10^{-13}$\\ \hline
1ES 2344+514 & BL Lac & 356.77 & 51.70 & 0.9518 & -10.1395 & 0.00 & 5.28$\cdot 10^{-13}$\\ \hline
IC443 & SNR & 94.18 & 22.53 & 0.9620 & -16.4154 & 0.00 & 3.63$\cdot 10^{-13}$\\ \hline
MGRO J2019+37 & PWN & 305.22 & 36.83 & 0.9784 & -17.6070 & 0.00 & 4.54$\cdot 10^{-13}$\\ \hline
\end{tabular}
\centering
\end{table*}

\begin{figure}
   	\centering
    \input{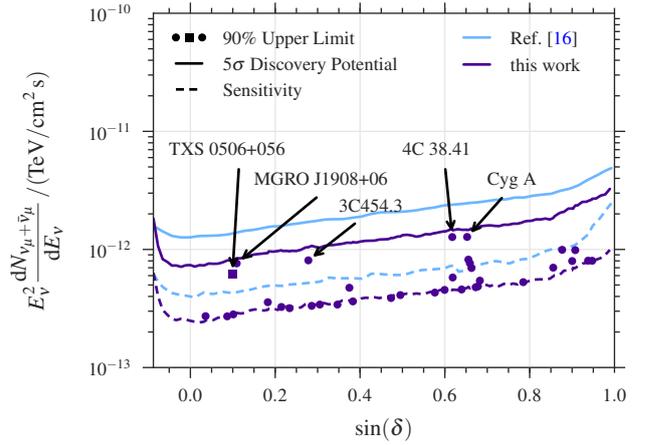}
    \caption{Sensitivity (dashed) and $5\sigma$ discovery potential (solid) of the flux normalization for an $E^{-2}$ source spectrum 
    as function of the $\sin\delta$. For comparison, the lines from~\cite{psPaper} are shown as well. 90\% CL Neyman upper limits on the flux normalization for sources in the \emph{a priori} and \emph{monitored} source list are shown as circles and squares, respectively.}
    \label{fig:limits_source_cat}
\end{figure}

\begin{figure*}[htbp]
    \centering
    \includegraphics[width=0.82\textwidth]{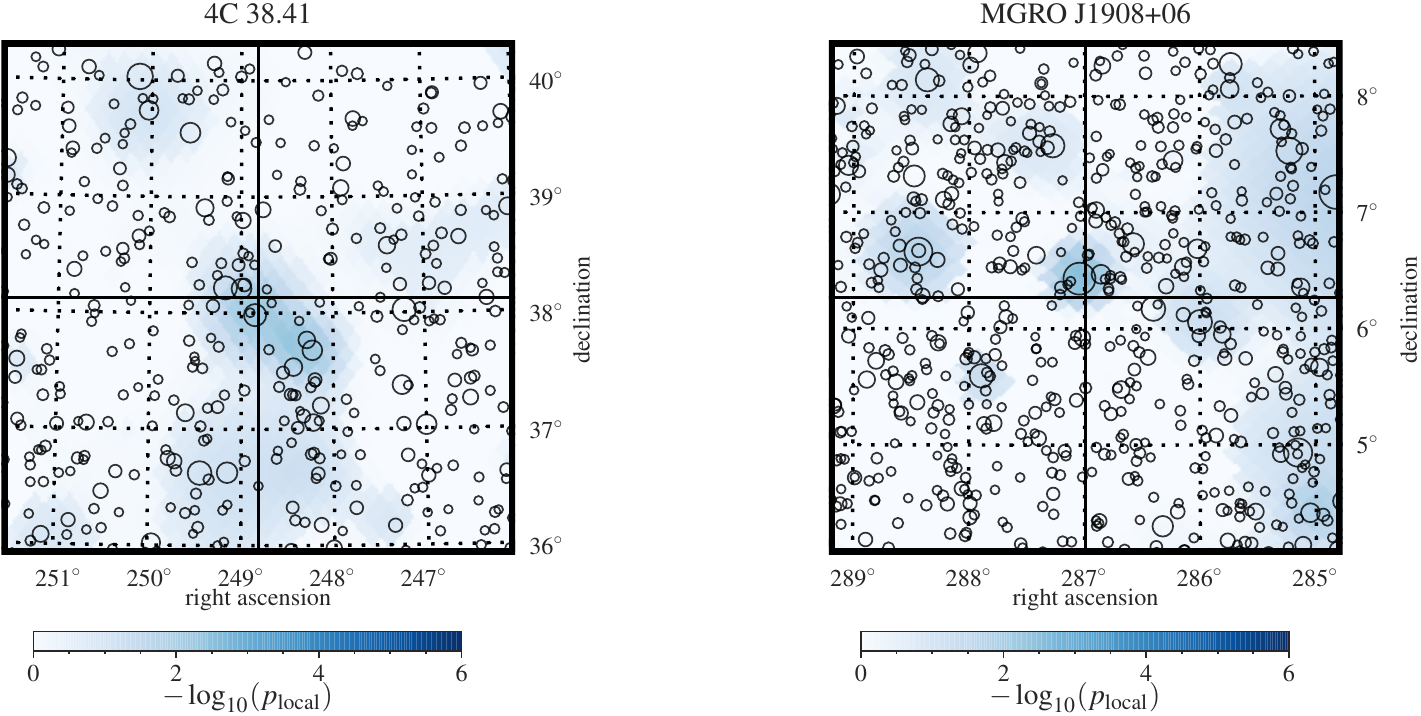}
    \caption{Local p-value landscapes around the source position of 4C 38.41 (left) and MGRO J1908+06 (right) in equatorial coordinates (J2000). 
        Neutrino event arrival directions are indicated by small circles where the area of the circle is proportional to the median $\log_{10}$ of neutrino energy assuming the diffuse best-fit spectrum.
        The p-value is evaluated at the point where the black lines cross.}
        \label{fig:4C_zoom} \label{fig:MGRO_zoom}
        \label{fig:catalog_landscapes}
\end{figure*}

\begin{figure*}[htbp]
    \centering
    \includegraphics[width=0.82\textwidth]{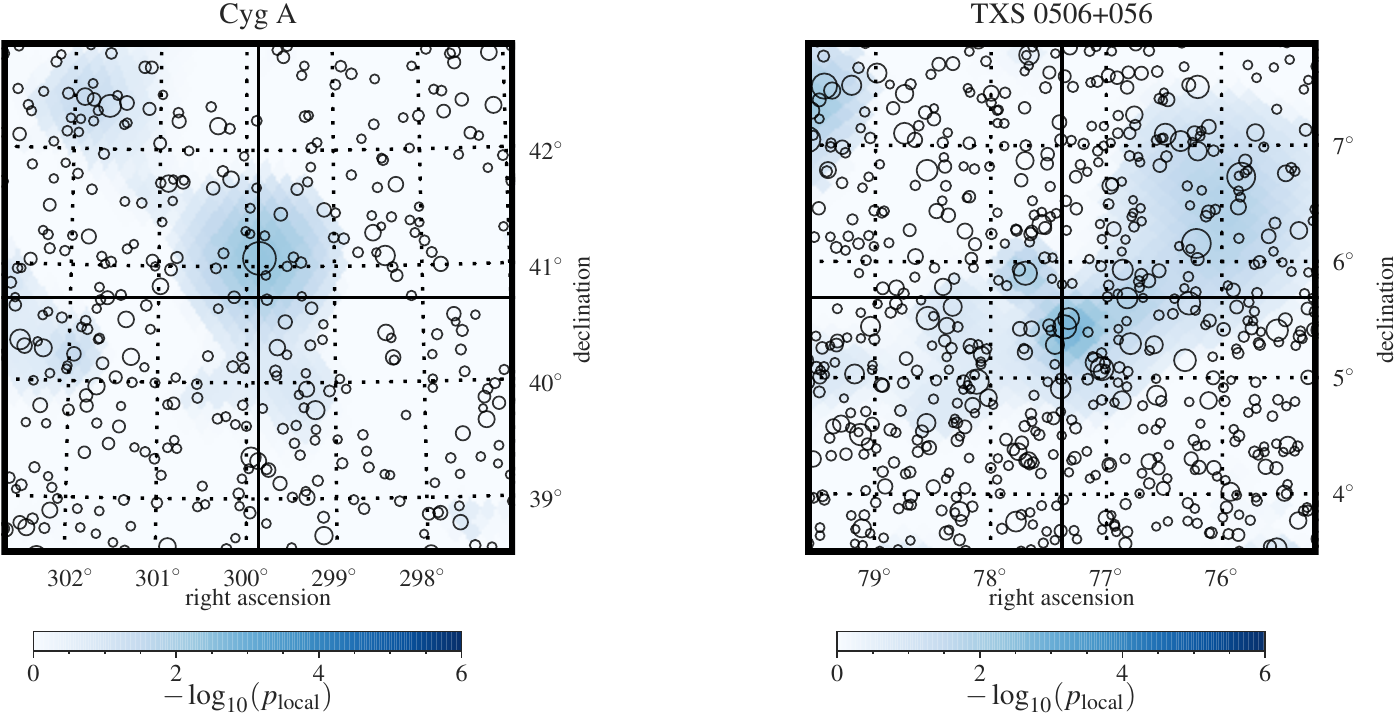}
    \caption{Local p-value landscapes around the source position of  Cyg A (left) and TXS 0506+056 (right) in equatorial coordinates (J2000).
        Neutrino event arrival directions are indicated by small circles where the area of the circle is proportional to the median $\log_{10}$ of neutrino energy assuming the diffuse best-fit spectrum.
        The p-value is evaluated at the point where the black lines cross.}
        \label{fig:Cyg_A}\label{fig:TXS_zoom}
        \label{fig:catalog_landscapes2}
\end{figure*}

The fit results of sources in the \emph{a priori} source list are given in Tab.~\ref{tab:results_source_cat}. 
The most significant source with a local p-value of 0.8\% is 4C 38.41, which is a flat spectrum radio quasar (FSRQ) at a redshift of $z=1.8$. Taking into account that 34 sources have been tested, a post-trial p-value of 23.7\% (20.3\%) is calculated from background pseudo-experiments which is compatible with background. 

As no significant source has been found, 90\% CL upper limits are calculated assuming an unbroken power law with spectral index of -2 using Neyman's method~\cite{neyman_upper_limit}. The 90\% CL upper limit flux is summarized in Tab.~\ref{tab:results_source_cat} and shown in Fig.~\ref{fig:limits_source_cat}.
In case of under-fluctuations, the limit was set to the sensitivity level of the analysis. Note that 90\% upper limits can exceed the discovery potential as long as the best-fit flux is below the discovery potential.

Interestingly, a total of
three  sources, 4C 38.41, MGRO J1908+06 and Cyg A, have a local p-value below or close to 1\%. 
The p-value landscapes and observed events around these three sources are shown in Fig.~\ref{fig:catalog_landscapes} and Fig.~\ref{fig:catalog_landscapes2}.

\subsection{Population test in the a piori source list}

\begin{figure}[htbp]
   	\centering
    \input{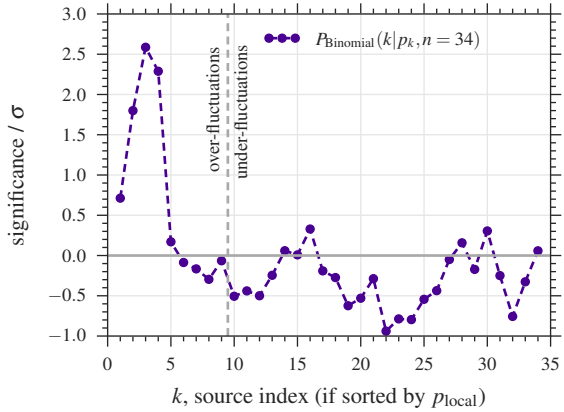}
    \caption{Local significance in Gaussian $\sigma$ for binomial combinations of the $k$ most significant sources in the \emph{a priori} source list. Sources with $\hat{n}_s>0$ and $\hat{n}_s=0$ can be separated by the dashed vertical line.}
    \label{fig:CPA_binomial}
\end{figure}

The most significant combination of p-values from the \emph{a priori} source list is given when combining the three most significant p-values, i.e. $k=3$, with $2.59\sigma$ as shown in Fig.~\ref{fig:CPA_binomial}. The comparison with background pseudo-experiments yields a trial-corrected p-value of 6.6\% (4.1\%) which is not significant. 

\subsection{Monitored source list}

\begin{table*}[htbp]
  \centering
  \caption{Results of the \emph{monitored} source list search. 
  The fitted spectral index $\hat{\gamma}$ is not given as it is effectively fixed by the introduced prior. We use the abbreviation BL Lac for BL Lacertae objects.\label{tab:monitored_sources}}
  \setlength\q{\dimexpr .1\textwidth -2\tabcolsep}
  \setlength\qq{\dimexpr .15\textwidth -2\tabcolsep}
  \setlength\qqq{\dimexpr .25\textwidth -2\tabcolsep}
  \begin{tabular}{p{\qq}p{\q}p{\q}p{\q}p{\q}p{\q}p{\q}c}
    Source & Type & $\alpha\,[\mathrm{deg}]$ & $\delta\,[\mathrm{deg}]$ & p-Value & $TS$ & $\hat{n}_s$ & $E^{2}\mathrm{d}N_{\nu_\mu+\bar{\nu}_\mu} / \mathrm{d}E\,[\mathrm{TeV}\,\mathrm{cm}^{-2}\,\mathrm{s}^{-1}]$\Tstrut\Bstrut \\ 
    \hline \hline 
    TXS 0506+056 & BL Lac & 77.38 & 5.69 & 0.0293 & 2.6475 & 7.87 & 6.19$\cdot 10^{-13}$\\ \hline 
  \end{tabular}
\end{table*}
The best-fit results for TXS 0506+056 in the \emph{monitored} source list are given in Tab.~\ref{tab:monitored_sources}. Note that the event selection ends in May 2017 and thus does not include the time of the alert ICECUBE-170922A \cite{icecube_collaboration_amon} that led to follow-up observations and the discovery of $\gamma$-ray emission from that blazar up to 400\,GeV. 
The data, however, include the earlier time-period of the observed neutrino flare. 
The local p-value here is found to be 2.93\%. This is less significant than the reported significance of the time-dependent flare in~\cite{IceCube:2018cha} but
is consistent with the reported
time-integrated significances
in \cite{IceCube:2018cha}, 
when taking into account that this analysis has a prior on the spectral index of the source flux and does not cover the same time-range as in~\cite{IceCube:2018cha}. 

The local p-value landscape around TXS 0506+056 is shown in Fig.~\ref{fig:catalog_landscapes2} together with the observed event directions of this sample.

\section{Implications on source populations}\label{sec:Kowalski}

\begin{figure}[htbp]
        \centering
        \input{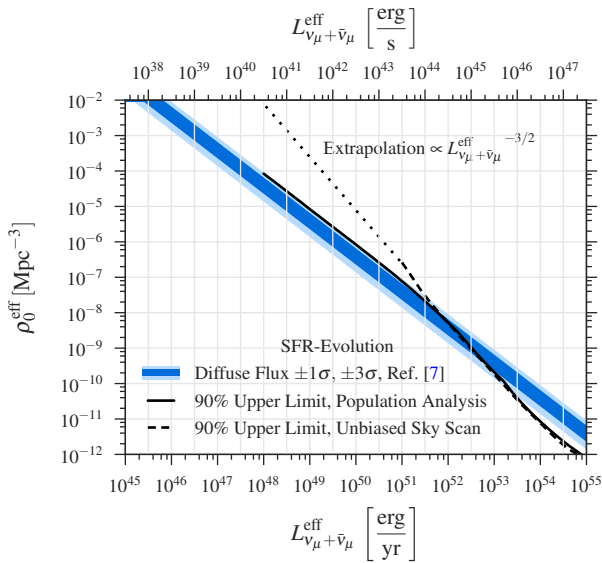}
        \caption{90\% CL upper limits on the effective muon-neutrino luminosity within the energy range $10^4\,\mathrm{GeV}$ -- $10^7\,\mathrm{GeV}$ at Earth and effective source density, derived from the hotspot population analysis and the  sky scan.}
        \label{fig:HPA_UL_Kowalski}
\end{figure}

The non-detection of a significant point-like source and the non-detection of a population of sources within the  sky scan is used to put constrains on realistic source populations.
In the following calculation, source populations are characterized by their effective $\nu_\mu + \bar{\nu}_\mu$ single-source luminosity $L_{\nu_\mu + \bar{\nu}_\mu}^\mathrm{eff}$ and their local source density $\rho_0^\mathrm{eff}$. 
Using the software tool FIRESONG\footnote{FIRst Extragalactic Simulation Of Neutrinos and Gamma-rays (FIRESONG), \url{https://github.com/ChrisCFTung/FIRESONG}}~\cite{taboada_constrains_2018}, the resulting source count distribution $\frac{\mathrm{d}N}{\mathrm{d}\Phi}$ as a function of the flux $\Phi$ for source populations are calculated for sources within $z<10$ and representations of this population are simulated. To calculate the source count distribution, FIRESONG takes the source density $\rho$, luminosity distribution, source evolution, cosmological parameters, the  energy range of the flux and the spectral index into account. Following Ref.~\cite{murase_constraining_2016}, sources are simulated with a log-normal distribution with median $L_{\nu_\mu + \bar{\nu}_\mu}^\mathrm{eff}$ and a width of 0.01 in $\log_{10}(L_{\nu_\mu + \bar{\nu}_\mu}^\mathrm{eff})$ which corresponds to a standard candle luminosity. The evolution of the sources was chosen to follow the parametrization of star formation rate from Hopkins and Beacom~\cite{hopkins_normalization_2006} assuming a flat universe with $\Omega_{M,0} =0.308$, $\Omega_{\lambda,0} = 0.692$ and $h = 0.678$~\cite{planck_collaboration_planck_2016}. The  energy range of the flux at Earth was chosen as $10^4\,\mathrm{GeV}$ -- $10^7\,\mathrm{GeV}$
to calculate the effective muon neutrino luminosities of sources.
 
Generating pseudo-experiments with signal components corresponding to the flux distribution obtained from FIRESONG, 90\% CL upper limits are calculated in the $\rho_0^\mathrm{eff}$ - $L_{\nu_\mu + \bar{\nu}_\mu}^\mathrm{eff}$ plane for various spectral indices assuming that background consists of atmospheric neutrinos only, as described in Section~\ref{sec:HPA_result}.
The 90\% CL upper limit is calculated based on the fact that the strongest source of a population does not give a p-value in the  sky scan that is larger than the observed one. The 90\% upper limits are shown as dashed lines in Fig.~\ref{fig:HPA_UL_Kowalski}. In addition, 90\% CL upper limits are calculated by comparing the largest excess measured with the population test in the  sky scan. These 90\% upper limits are shown as solid lines in Fig.~\ref{fig:HPA_UL_Kowalski}. Populations that are compatible at the $1\sigma$ and $3\sigma$ level with the diffuse flux measured in~\cite{MUONICRC2017} are shown as blue shaded band.
90\% CL upper limits have been calculated assuming an $E^{-2}$ power-law flux. The same has been performed for an $E^{-2.19}$ power-law flux, which is the diffuse best-fit for this sample (this result can be found in the supplementary material). The computation of upper limits becomes very computing-intensive for large source densities. Therefore, the computation of the upper limits, resulting from the  sky scan, are extrapolated to larger source densities (indicated by dotted line in Fig.~\ref{fig:HPA_UL_Kowalski}). 
It can be seen that for large effective source densities and small effective luminosities, the limit resulting from the population analysis goes $\propto 1 / L_{\nu_\mu + \bar{\nu}_\mu}^\mathrm{eff}$ which is the same scaling as one would expect from a diffuse flux. Indeed it is found that an excess of diffuse high-energy events, i.e. sources from which only one neutrino are detected, leads to a p-value excess in the population analysis. This is a result of taking the energy of the event into account in the likelihood. Limits from the hottest spot in the  sky scan are a bit stronger for large effective luminosities while upper limits from the population test become stronger at about $L_{\nu_\mu + \bar{\nu}_\mu}^\mathrm{eff} \sim 10^{52} \frac{\mathrm{erg}}{\mathrm{yr}}$. 

\section{Implications for individual source models}\label{sec:source_models}

\begin{table*}[htbp]
    \centering
    \caption{Model rejection factors for source models in the source catalog. Given are source type, model reference, central energy range that contributes 90\% to sensitivity, MRF sensitivity and MRF at 90\% CL.}
    \label{tab:MRF}
    \begin{tabular}{lllll}
        Type & Source Model & $\log_{10}(E/\mathrm{GeV})$ & sensitivity & 90\% UL \Tstrut\Bstrut \\ 
        \hline \hline 
        Crab & Amato \textit{et. al} \cite{amatosignatures2003} $\Gamma=10^4$ & 1.5 - 9.0 & 23.38 & 31.47  \\ \hline 
        & Amato \textit{et. al} \cite{amatosignatures2003} $\Gamma=10^5$ & 3.0 - 4.5 & 0.79 & 1.14  \\ \hline 
        & Amato \textit{et. al} \cite{amatosignatures2003} $\Gamma=10^6$ & 4.0 - 5.5 & 0.16 & 0.21  \\ \hline 
        & Amato \textit{et. al} \cite{amatosignatures2003} $\Gamma=10^7$ & 4.5 - 6.0 & 0.32 & 0.40  \\ \hline 
        & Kappes \textit{et. al} \cite{kappespotential2007} & 2.5 - 4.5 & 1.06 & 1.47 \\ \hline 
        Blazar & 3C273, Reimer \cite{reimerphotonneutrino}  & 6.0 - 8.5 & 0.39 & 0.42 \\ \hline 
        & 3C454.3, Reimer \cite{reimerphotonneutrino} & 6.0 - 8.0 & 2.80 & 5.42  \\ \hline 
        & Mrk421, Petropoulou \textit{et. al} \cite{petropoulouphotohadronic2015}  & 5.5 - 7.0 & 0.36 & 0.43 \\ \hline 
        SNR & G40.5-0.5, Mandelartz \textit{et. al} \cite{mandelartzprediction2015} & 3.5 - 5.5 & 1.45 & 4.57  \\ \hline 
    \end{tabular}
\end{table*}

In Section~\ref{sec:source_list}, constraints on source fluxes assuming $\mathrm{d}N / \mathrm{d}E_\nu \propto E_\nu^{-2}$ have been calculated. However, more specific neutrino flux models can be obtained using $\gamma$-ray data. In pion decays, both neutrinos and $\gamma$-rays are produced. Thus $\gamma$-ray data can be used to construct models for neutrino emission under certain assumptions. Here, models for sources of the \emph{a priori} source list are tested. 
For each model, the Model Rejection Factor (MRF) is calculated which is the ratio between the predicted flux and the 90\% CL upper limit. 
In addition, the expected experimental result in the case of pure background is also calculated giving the MRF sensitivity. The energy range that contributes 90\% to the sensitivity has been calculated by folding the differential discovery potential at the source position (similar to Fig.~\ref{fig:differential_sensitivity_disc_pot_E2}) with the flux prediction. Models for which the MRF sensitivity is larger than 10 are not discussed here.

The first source tested is the Crab Nebula, which is a Pulsar Wind Nebula (PWN) and the brightest source in TeV $\gamma$-rays. Despite the common understanding that the emission from PWNe is of leptonic nature, see e.g.~\cite{aleksic_measurement_2015}, neutrinos can be produced by subdominant hadronic emission. 
Predictions for neutrino fluxes from the Crab Nebula are proposed, e.g. by Amato \textit{et. al}~\cite{amatosignatures2003} and  Kappes  \textit{et. al}~\cite{kappespotential2007}. 
The prediction by Amato \textit{et. al} assumes pion production is dominated by p-p interactions and the target density is given by $n_t = 10\mu M_{N_\odot} R_\mathrm{pc}^{-3}\, \mathrm{cm}^{-3}$ with $M_{N_\odot}$ the mass of the supernova ejecta in units of solar masses. Moreover, $R_\mathrm{pc}$ is the radius of the supernova in units of $\mathrm{pc}$ and $\mu$ is an unknown factor of the order of $1\leq \mu \leq 20$ that takes into account e.g. the intensity and structures of magnetic fields within the PWN. Here $\mu=20$ and a proton luminosity of 60\% of the total PWN luminosity for Lorentz factors of $\Gamma = 10^4,\, 10^5,\,10^6,\,10^7$ are used to provide a result that is model-independent and complementary to~\cite{amatosignatures2003}. 
The model prediction by Kappes~\textit{et al.},
assumes a dominant production
of $\gamma$-rays of the HESS $\gamma$-ray spectrum~\cite{aharonian_observations_2006}
by p-p interactions.

\begin{figure}[htbp]
        \centering
        \input{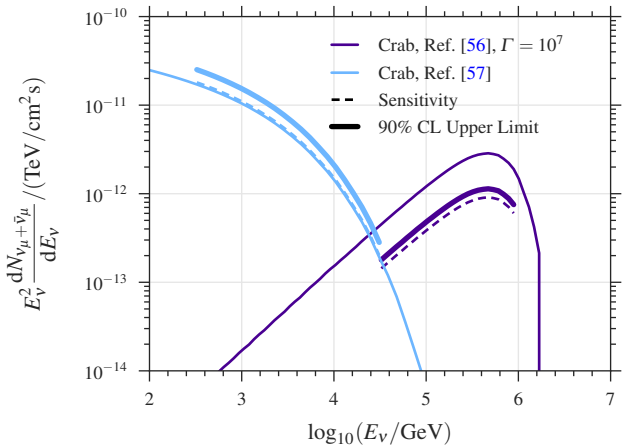}
        \caption{Differential source flux for the Crab Nebula. Solid lines show the model prediction, thick lines give the 90\% CL upper limit and the dashed lines indicate the sensitivity flux.  90\% CL upper limit and sensitivity are shown in the energy range that contributes 90\% to the sensitivity.}
        \label{fig:MRF_Crab}
\end{figure}

The model predictions, sensitivity and 90\% CL upper limit are shown in Fig.~\ref{fig:MRF_Crab} and are listed in Tab.~\ref{tab:MRF}. Sensitivity and upper limits are shown for the central energy range that contributes 90\% to the sensitivity.

For the model of Kappes \textit{et al.}, the sensitivity is very close to the model prediction while for Amato \textit{et al.} with $\Gamma=10^7$, the sensitivity is a factor of three lower than the prediction. The 90\% CL upper limits are listed in Tab.~\ref{tab:MRF}. They are slightly higher but still constrain the models by Amato \textit{et al.}

\begin{figure}[htp]
        \centering
        \input{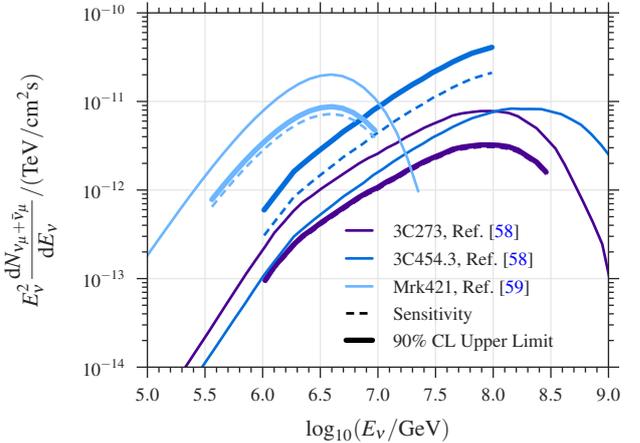}
        \caption{Differential source flux for 3C273, 3C454.3 and Mrk 421. Solid lines show the model prediction, thick lines give the 90\% CL upper limit and dashed lines indicate the sensitivity flux. 90\% CL upper limit and sensitivity are shown in the energy range that contributes 90\% to the sensitivity.}
        \label{fig:MRF_Blazars}
\end{figure}

Another very interesting class of sources are active galactic nuclei (AGN). Here, the models being tested come from Ref.~\cite{petropoulouphotohadronic2015} for Mrk 421, a BL Lacertae object (BL Lac) that was found in spatial and energetic agreement with a high-energy starting event and from Ref.~\cite{reimerphotonneutrino} for 3C273 and 3C454.3 which are flat spectrum radio quasars (FSRQ). The models, sensitivities and 90\% CL upper limits are shown in Fig.~\ref{fig:MRF_Blazars}
 and the MRF are listed in Tab.~\ref{tab:MRF}. 

The sensitivities for 3C273 and Mrk 421 are well below the model prediction and the 90\% CL upper limits are at about 40\% of the model flux. For 3C454.3, the sensitivity is a factor 2.8 above the model prediction. Since 3C454.3 is one of the few sources with a local p-value below $\sim2.5\%$, the 90\% CL upper limit is much larger.

\begin{figure}[htp]
        \centering
        \input{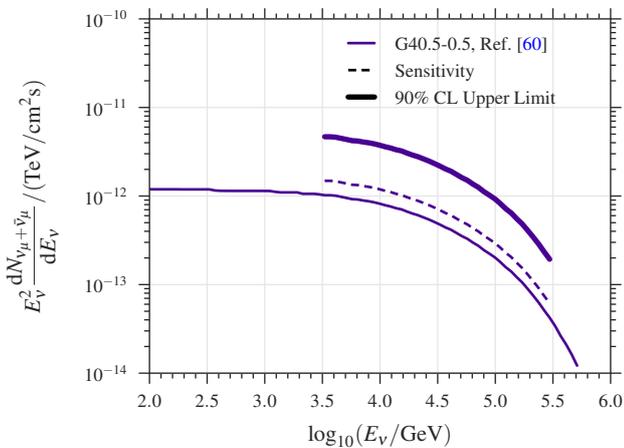}
        \caption{Differential source flux for SNR G40.5-0.5. The solid line gives the model prediction, the thick line gives the 90\% CL upper limit and the dashed line indicates the sensitivity flux. The 90\% CL upper limit and sensitivity are shown in the energy range that contributes 90\% to the sensitivity. G40.5-0.5 is associated with MGRO J1908+06.}
        \label{fig:MRF_SNR}
\end{figure}

Another tested model was derived for the source G40.5-0.5 which is a galactic supernova remnant~\cite{mandelartzprediction2015}. This supernova remnant can be associated with the TeV source MGRO J1908+06 which is the second most significant source in the \emph{a priori} source catalog, although the association of G40.5-0.5 with MGRO J1908+06 is not distinct~\cite{aliu_investigating_2014}. In addition, the pulsar wind nebula powered by PSR J1907+0602 may contribute to the TeV emission of the MGRO J1908+06 region. However, here the tested model for the SNR G40.5-0.5 is adapted from Ref.~\cite{mandelartzprediction2015}. The model, sensitivity and 90\% CL upper limit are shown in Fig.~\ref{fig:MRF_SNR} and are listed in Tab.~\ref{tab:MRF}.

The sensitivity of this analysis
is a factor 1.4 above the model prediction and not yet sensitive to this model. As MGRO J1908+06 is the second most significant source in the catalog, with a local p-value of $<1\%$, the  upper limit lies nearly a factor of five above the model prediction.

\section{Conclusions}\label{sec:conclusions}

Eight years of IceCube data have been analyzed for a time-independent clustering of through-going muon neutrinos using an unbinned likelihood method. 
The analysis includes a full sky search of the Northern hemisphere down to a declination of $-3^\circ $  for a significant hot spot as well as an  analysis of a possible cumulative excess of a population of weak sources. Furthermore, source-candidates from an \emph{a priori} catalog and a catalog of monitored sources are tested individually and again for a cumulative excess.

The analysis method has been optimized for the observed energy spectrum of high-energy astro-physical muon neutrinos \cite{diffuse_muon_paper} and a number of improvements with respect to the previously published search \cite{psPaper} have been incorporated. By implementing these improvements, a sensitivity increase of about 35\% has been achieved.

No significant source was found in the full-sky  scan of the Northern hemisphere and the search for significant neutrino emission from objects on a \emph{a priori} source list results in a post-trial p-value of 23.7\% (20.3\%), compatible with background. Also the tests for populations of sub-threshold sources revealed no significant excess.

Three sources on the \emph{a priori} source-list,
4C 38.41, MGRO J1908+06 and Cyg A,
have pre-trial p-values of only about 1\%. However, these excesses are not significant. 
The source TXS 0506+056 
in the catalog of monitored sources 
has a p-value of 2.9 \%.
This is consistent with the time-integrated p-value in \cite{IceCube:2018cha} for the assumed prior on the spectral index. 

Based on these results, the most stringent limits on high-energy neutrino emission from point-like sources are obtained. In addition, models for neutrino emission from specific sources are tested. The model \cite{amatosignatures2003} for the Crab Nebula is excluded for $\Gamma \ge 10^6 $ as well as the predictions for 3C273 \cite{reimerphotonneutrino} and Mrk 421 \cite{petropoulouphotohadronic2015}.
In addition to these specific models, an exclusion of source populations as a function of local source density and single-source luminosity are derived by calculating the source count distribution for a realistic cosmological evolution model. 

\begin{acknowledgements}
The IceCube collaboration acknowledges the significant contributions to this manuscript from Ren\'e Reimann.
The authors gratefully acknowledge the support from the following agencies and institutions: 
USA -- U.S. National Science Foundation-Office of Polar Programs,
U.S. National Science Foundation-Physics Division,
Wisconsin Alumni Research Foundation,
Center for High Throughput Computing (CHTC) at the University of Wisconsin-Madison,
Open Science Grid (OSG),
Extreme Science and Engineering Discovery Environment (XSEDE),
U.S. Department of Energy-National Energy Research Scientific Computing Center,
Particle astrophysics research computing center at the University of Maryland,
Institute for Cyber-Enabled Research at Michigan State University,
and Astroparticle physics computational facility at Marquette University;
Belgium -- Funds for Scientific Research (FRS-FNRS and FWO),
FWO Odysseus and Big Science programmes,
and Belgian Federal Science Policy Office (Belspo);
Germany -- Bundesministerium f\"ur Bildung und Forschung (BMBF),
Deutsche Forschungsgemeinschaft (DFG),
Helmholtz Alliance for Astroparticle Physics (HAP),
Initiative and Networking Fund of the Helmholtz Association,
Deutsches Elektronen Synchrotron (DESY),
and High Performance Computing cluster of the RWTH Aachen;
Sweden -- Swedish Research Council,
Swedish Polar Research Secretariat,
Swedish National Infrastructure for Computing (SNIC),
and Knut and Alice Wallenberg Foundation;
Australia -- Australian Research Council;
Canada -- Natural Sciences and Engineering Research Council of Canada,
Calcul Qu\'ebec, Compute Ontario, Canada Foundation for Innovation, WestGrid, and Compute Canada;
Denmark -- Villum Fonden, Danish National Research Foundation (DNRF), Carlsberg Foundation;
New Zealand -- Marsden Fund;
Japan -- Japan Society for Promotion of Science (JSPS)
and Institute for Global Prominent Research (IGPR) of Chiba University;
Korea -- National Research Foundation of Korea (NRF);
Switzerland -- Swiss National Science Foundation (SNSF).

\end{acknowledgements}
%%%%%%%%%%%%%%%%%%%%%%%%%%%%%%%%%%%%%%%%% bibliography %%%%%%%%%%%%%%%%%%%%%%%%%%%%%%%%%%%%%%%%%%

%\bibliographystyle{apsrev4-1}
%\bibliography{reference}

\begin{thebibliography}{61}%
\makeatletter
\providecommand \@ifxundefined [1]{%
 \@ifx{#1\undefined}
}%
\providecommand \@ifnum [1]{%
 \ifnum #1\expandafter \@firstoftwo
 \else \expandafter \@secondoftwo
 \fi
}%
\providecommand \@ifx [1]{%
 \ifx #1\expandafter \@firstoftwo
 \else \expandafter \@secondoftwo
 \fi
}%
\providecommand \natexlab [1]{#1}%
\providecommand \enquote  [1]{``#1''}%
\providecommand \bibnamefont  [1]{#1}%
\providecommand \bibfnamefont [1]{#1}%
\providecommand \citenamefont [1]{#1}%
\providecommand \href@noop [0]{\@secondoftwo}%
\providecommand \href [0]{\begingroup \@sanitize@url \@href}%
\providecommand \@href[1]{\@@startlink{#1}\@@href}%
\providecommand \@@href[1]{\endgroup#1\@@endlink}%
\providecommand \@sanitize@url [0]{\catcode `\\12\catcode `\$12\catcode
  `\&12\catcode `\#12\catcode `\^12\catcode `\_12\catcode `\%12\relax}%
\providecommand \@@startlink[1]{}%
\providecommand \@@endlink[0]{}%
\providecommand \url  [0]{\begingroup\@sanitize@url \@url }%
\providecommand \@url [1]{\endgroup\@href {#1}{\urlprefix }}%
\providecommand \urlprefix  [0]{URL }%
\providecommand \Eprint [0]{\href }%
\providecommand \doibase [0]{http://dx.doi.org/}%
\providecommand \selectlanguage [0]{\@gobble}%
\providecommand \bibinfo  [0]{\@secondoftwo}%
\providecommand \bibfield  [0]{\@secondoftwo}%
\providecommand \translation [1]{[#1]}%
\providecommand \BibitemOpen [0]{}%
\providecommand \bibitemStop [0]{}%
\providecommand \bibitemNoStop [0]{.\EOS\space}%
\providecommand \EOS [0]{\spacefactor3000\relax}%
\providecommand \BibitemShut  [1]{\csname bibitem#1\endcsname}%
\let\auto@bib@innerbib\@empty
%</preamble>
\bibitem [{\citenamefont {Gaisser}\ \emph {et~al.}(1995)\citenamefont
  {Gaisser}, \citenamefont {Halzen},\ and\ \citenamefont
  {Stanev}}]{Gaisser:1994yf}%
  \BibitemOpen
  \bibfield  {author} {\bibinfo {author} {\bibfnamefont {T.~K.}\ \bibnamefont
  {Gaisser}}, \bibinfo {author} {\bibfnamefont {F.}~\bibnamefont {Halzen}}, \
  and\ \bibinfo {author} {\bibfnamefont {T.}~\bibnamefont {Stanev}},\ }\href
  {\doibase 10.1016/0370-1573(95)00003-Y} {\bibfield  {journal} {\bibinfo
  {journal} {Phys. Rept.}\ }\textbf {\bibinfo {volume} {258}},\ \bibinfo
  {pages} {173} (\bibinfo {year} {1995})},\ \bibinfo {note} {[Erratum: Phys.
  Rept.271,355(1996)]},\ \Eprint {http://arxiv.org/abs/hep-ph/9410384}
  {arXiv:hep-ph/9410384} \BibitemShut {NoStop}%
%%CITATION = HEP-PH/9410384;%%
\bibitem [{\citenamefont {Athar}\ \emph {et~al.}(2006)\citenamefont {Athar},
  \citenamefont {Kim},\ and\ \citenamefont {Lee}}]{Athar:2005wg}%
  \BibitemOpen
  \bibfield  {author} {\bibinfo {author} {\bibfnamefont {H.}~\bibnamefont
  {Athar}}, \bibinfo {author} {\bibfnamefont {C.~S.}\ \bibnamefont {Kim}}, \
  and\ \bibinfo {author} {\bibfnamefont {J.}~\bibnamefont {Lee}},\ }\href
  {\doibase 10.1142/S021773230602038X} {\bibfield  {journal} {\bibinfo
  {journal} {Mod. Phys. Lett.}\ }\textbf {\bibinfo {volume} {A21}},\ \bibinfo
  {pages} {1049} (\bibinfo {year} {2006})},\ \Eprint
  {http://arxiv.org/abs/hep-ph/0505017} {arXiv:hep-ph/0505017} \BibitemShut
  {NoStop}%
%%CITATION = HEP-PH/0505017;%%
\bibitem [{\citenamefont {Aartsen}\ \emph
  {et~al.}(2013{\natexlab{a}})\citenamefont {Aartsen} \emph
  {et~al.}}]{hese_paper}%
  \BibitemOpen
  \bibfield  {author} {\bibinfo {author} {\bibfnamefont {M.~G.}\ \bibnamefont
  {Aartsen}} \emph {et~al.} (\bibinfo {collaboration} {IceCube
  Collaboration}),\ }\href {\doibase 10.1126/science.1242856} {\bibfield
  {journal} {\bibinfo  {journal} {Science}\ }\textbf {\bibinfo {volume}
  {342}},\ \bibinfo {pages} {1242856} (\bibinfo {year} {2013}{\natexlab{a}})},\
  \Eprint {http://arxiv.org/abs/1311.5238} {arXiv:1311.5238} \BibitemShut
  {NoStop}%
\bibitem [{\citenamefont {Kopper}(2017)}]{HESEICRC2017}%
  \BibitemOpen
  \bibfield  {author} {\bibinfo {author} {\bibfnamefont {C.}~\bibnamefont
  {Kopper}} (\bibinfo {collaboration} {{IceCube Collaboration}}),\ }\href
  {https://pos.sissa.it/301/981/} {\bibfield  {journal} {\bibinfo  {journal}
  {PoS}\ }\textbf {\bibinfo {volume} {ICRC2017}},\ \bibinfo {pages} {981}
  (\bibinfo {year} {2017})},\ \Eprint {http://arxiv.org/abs/1710.01191}
  {arXiv:1710.01191} \BibitemShut {NoStop}%
\bibitem [{\citenamefont {Aartsen}\ \emph
  {et~al.}(2015{\natexlab{a}})\citenamefont {Aartsen} \emph
  {et~al.}}]{CHRISPRL}%
  \BibitemOpen
  \bibfield  {author} {\bibinfo {author} {\bibfnamefont {M.~G.}\ \bibnamefont
  {Aartsen}} \emph {et~al.} (\bibinfo {collaboration} {IceCube
  Collaboration}),\ }\href {\doibase 10.1103/PhysRevLett.115.081102} {\bibfield
   {journal} {\bibinfo  {journal} {Phys. Rev. Lett.}\ }\textbf {\bibinfo
  {volume} {115}},\ \bibinfo {pages} {081102} (\bibinfo {year}
  {2015}{\natexlab{a}})},\ \Eprint {http://arxiv.org/abs/1507.04005}
  {arXiv:1507.04005} \BibitemShut {NoStop}%
%%CITATION = ARXIV:1507.04005;%%
\bibitem [{\citenamefont {Aartsen}\ \emph
  {et~al.}(2016{\natexlab{a}})\citenamefont {Aartsen} \emph
  {et~al.}}]{diffuse_muon_paper}%
  \BibitemOpen
  \bibfield  {author} {\bibinfo {author} {\bibfnamefont {M.~G.}\ \bibnamefont
  {Aartsen}} \emph {et~al.} (\bibinfo {collaboration} {IceCube
  Collaboration}),\ }\href {\doibase 10.3847/0004-637X/833/1/3} {\bibfield
  {journal} {\bibinfo  {journal} {Astrophys. J.}\ }\textbf {\bibinfo {volume}
  {833}},\ \bibinfo {pages} {3} (\bibinfo {year} {2016}{\natexlab{a}})},\
  \Eprint {http://arxiv.org/abs/1607.08006} {arXiv:1607.08006} \BibitemShut
  {NoStop}%
\bibitem [{\citenamefont {Haack}(2017)}]{MUONICRC2017}%
  \BibitemOpen
  \bibfield  {author} {\bibinfo {author} {\bibfnamefont {C.}~\bibnamefont
  {Haack}} (\bibinfo {collaboration} {{IceCube Collaboration}}),\ }\href
  {https://pos.sissa.it/301/1005/} {\bibfield  {journal} {\bibinfo  {journal}
  {PoS}\ }\textbf {\bibinfo {volume} {ICRC2017}},\ \bibinfo {pages} {1005}
  (\bibinfo {year} {2017})},\ \Eprint {http://arxiv.org/abs/1710.01191}
  {arXiv:1710.01191} \BibitemShut {NoStop}%
\bibitem [{\citenamefont {Aartsen}\ \emph
  {et~al.}(2018{\natexlab{a}})\citenamefont {Aartsen} \emph
  {et~al.}}]{IceCube:2018cha}%
  \BibitemOpen
  \bibfield  {author} {\bibinfo {author} {\bibfnamefont {M.~G.}\ \bibnamefont
  {Aartsen}} \emph {et~al.} (\bibinfo {collaboration} {IceCube
  Collaboration}),\ }\href {\doibase 10.1126/science.aat2890} {\bibfield
  {journal} {\bibinfo  {journal} {Science}\ }\textbf {\bibinfo {volume}
  {361}},\ \bibinfo {pages} {147} (\bibinfo {year} {2018}{\natexlab{a}})},\
  \Eprint {http://arxiv.org/abs/1807.08794} {arXiv:1807.08794} \BibitemShut
  {NoStop}%
%%CITATION = ARXIV:1807.08794;%%
\bibitem [{\citenamefont {Aartsen}\ \emph
  {et~al.}(2018{\natexlab{b}})\citenamefont {Aartsen} \emph
  {et~al.}}]{IceCube:2018dnn}%
  \BibitemOpen
  \bibfield  {author} {\bibinfo {author} {\bibfnamefont {M.~G.}\ \bibnamefont
  {Aartsen}} \emph {et~al.} (\bibinfo {collaboration} {Liverpool Telescope,
  MAGIC, H.E.S.S., AGILE, Kiso, VLA/17B-403, INTEGRAL, Kapteyn, Subaru, HAWC,
  Fermi-LAT, ASAS-SN, VERITAS, Kanata, IceCube, Swift NuSTAR}),\ }\href
  {\doibase 10.1126/science.aat1378} {\bibfield  {journal} {\bibinfo  {journal}
  {Science}\ }\textbf {\bibinfo {volume} {361}},\ \bibinfo {pages} {eaat1378}
  (\bibinfo {year} {2018}{\natexlab{b}})},\ \Eprint
  {http://arxiv.org/abs/1807.08816} {arXiv:1807.08816} \BibitemShut {NoStop}%
%%CITATION = ARXIV:1807.08816;%%
\bibitem [{\citenamefont {Aartsen}\ \emph
  {et~al.}(2017{\natexlab{a}})\citenamefont {Aartsen} \emph
  {et~al.}}]{aartsen_contribution_2017}%
  \BibitemOpen
  \bibfield  {author} {\bibinfo {author} {\bibfnamefont {M.~G.}\ \bibnamefont
  {Aartsen}} \emph {et~al.} (\bibinfo {collaboration} {IceCube
  Collaboration}),\ }\href {\doibase 10/f9r5p4} {\bibfield  {journal} {\bibinfo
   {journal} {{Astrophys. J.}}\ }\textbf {\bibinfo {volume} {835}},\ \bibinfo
  {pages} {45} (\bibinfo {year} {2017}{\natexlab{a}})},\ \Eprint
  {http://arxiv.org/abs/1611.03874} {arXiv:1611.03874} \BibitemShut {NoStop}%
\bibitem [{\citenamefont {Aartsen}\ \emph
  {et~al.}(2015{\natexlab{b}})\citenamefont {Aartsen} \emph
  {et~al.}}]{aartsen_search_2015}%
  \BibitemOpen
  \bibfield  {author} {\bibinfo {author} {\bibfnamefont {M.~G.}\ \bibnamefont
  {Aartsen}} \emph {et~al.} (\bibinfo {collaboration} {IceCube
  Collaboration}),\ }\href {\doibase 10/f3m3zm} {\bibfield  {journal} {\bibinfo
   {journal} {Astrophys. J.}\ }\textbf {\bibinfo {volume} {805}},\ \bibinfo
  {pages} {L5} (\bibinfo {year} {2015}{\natexlab{b}})},\ \Eprint
  {http://arxiv.org/abs/1412.6510} {arXiv:1412.6510} \BibitemShut {NoStop}%
\bibitem [{\citenamefont {Aartsen}\ \emph
  {et~al.}(2015{\natexlab{c}})\citenamefont {Aartsen} \emph
  {et~al.}}]{aartsen_searches_2015}%
  \BibitemOpen
  \bibfield  {author} {\bibinfo {author} {\bibfnamefont {M.~G.}\ \bibnamefont
  {Aartsen}} \emph {et~al.} (\bibinfo {collaboration} {IceCube
  Collaboration}),\ }\href {\doibase 10/f3nskj} {\bibfield  {journal} {\bibinfo
   {journal} {Astrophys. J.}\ }\textbf {\bibinfo {volume} {807}},\ \bibinfo
  {pages} {46} (\bibinfo {year} {2015}{\natexlab{c}})},\ \Eprint
  {http://arxiv.org/abs/1503.00598} {arXiv:1503.00598} \BibitemShut {NoStop}%
\bibitem [{\citenamefont {Adri\'{a}n-Mart\'{i}nez}\ \emph
  {et~al.}(2016)\citenamefont {Adri\'{a}n-Mart\'{i}nez} \emph
  {et~al.}}]{adrian-martinez_first_2016}%
  \BibitemOpen
  \bibfield  {author} {\bibinfo {author} {\bibfnamefont {S.}~\bibnamefont
  {Adri\'{a}n-Mart\'{i}nez}} \emph {et~al.} (\bibinfo {collaboration} {{ANTARES
  Collaboration} \& {IceCube Collaboration}}),\ }\href {\doibase 10/f3rzd9}
  {\bibfield  {journal} {\bibinfo  {journal} {Astrophys. J.}\ }\textbf
  {\bibinfo {volume} {823}},\ \bibinfo {pages} {65} (\bibinfo {year} {2016})},\
  \Eprint {http://arxiv.org/abs/1511.02149} {arXiv:1511.02149} \BibitemShut
  {NoStop}%
\bibitem [{\citenamefont {Aartsen}\ \emph
  {et~al.}(2016{\natexlab{b}})\citenamefont {Aartsen} \emph
  {et~al.}}]{aartsen_lowering_2016}%
  \BibitemOpen
  \bibfield  {author} {\bibinfo {author} {\bibfnamefont {M.~G.}\ \bibnamefont
  {Aartsen}} \emph {et~al.} (\bibinfo {collaboration} {IceCube
  Collaboration}),\ }\href {\doibase 10/f3rj3z} {\bibfield  {journal} {\bibinfo
   {journal} {{Astrophys. J.}}\ }\textbf {\bibinfo {volume} {824}},\ \bibinfo
  {pages} {L28} (\bibinfo {year} {2016}{\natexlab{b}})},\ \Eprint
  {http://arxiv.org/abs/1605.00163} {arXiv:1605.00163} \BibitemShut {NoStop}%
\bibitem [{\citenamefont {Aartsen}\ \emph
  {et~al.}(2016{\natexlab{c}})\citenamefont {Aartsen} \emph
  {et~al.}}]{a_very_2016}%
  \BibitemOpen
  \bibfield  {author} {\bibinfo {author} {\bibfnamefont {M.~G.}\ \bibnamefont
  {Aartsen}} \emph {et~al.} (\bibinfo {collaboration} {IceCube
  Collaboration}),\ }\href {\doibase 10/f97rfn} {\bibfield  {journal} {\bibinfo
   {journal} {J. Inst.}\ }\textbf {\bibinfo {volume} {11}},\ \bibinfo {pages}
  {P11009} (\bibinfo {year} {2016}{\natexlab{c}})},\ \Eprint
  {http://arxiv.org/abs/1610.01814} {arXiv:1610.01814} \BibitemShut {NoStop}%
\bibitem [{\citenamefont {Aartsen}\ \emph
  {et~al.}(2017{\natexlab{b}})\citenamefont {Aartsen} \emph
  {et~al.}}]{psPaper}%
  \BibitemOpen
  \bibfield  {author} {\bibinfo {author} {\bibfnamefont {M.~G.}\ \bibnamefont
  {Aartsen}} \emph {et~al.} (\bibinfo {collaboration} {IceCube
  Collaboration}),\ }\href {\doibase 10.3847/1538-4357/835/2/151} {\bibfield
  {journal} {\bibinfo  {journal} {Astrophys. J.}\ }\textbf {\bibinfo {volume}
  {835}},\ \bibinfo {pages} {151} (\bibinfo {year} {2017}{\natexlab{b}})},\
  \Eprint {http://arxiv.org/abs/1609.04981} {arXiv:1609.04981} \BibitemShut
  {NoStop}%
\bibitem [{\citenamefont {Aartsen}\ \emph
  {et~al.}(2017{\natexlab{c}})\citenamefont {Aartsen} \emph
  {et~al.}}]{aartsen_extending_2017}%
  \BibitemOpen
  \bibfield  {author} {\bibinfo {author} {\bibfnamefont {M.~G.}\ \bibnamefont
  {Aartsen}} \emph {et~al.} (\bibinfo {collaboration} {IceCube
  Collaboration}),\ }\href {\doibase 10/gbvxz6} {\bibfield  {journal} {\bibinfo
   {journal} {{Astrophys. J.}}\ }\textbf {\bibinfo {volume} {843}},\ \bibinfo
  {pages} {112} (\bibinfo {year} {2017}{\natexlab{c}})},\ \Eprint
  {http://arxiv.org/abs/1702.06868} {arXiv:1702.06868} \BibitemShut {NoStop}%
\bibitem [{\citenamefont {Aartsen}\ \emph
  {et~al.}(2017{\natexlab{d}})\citenamefont {Aartsen} \emph
  {et~al.}}]{aartsen_multiwavelength_2017}%
  \BibitemOpen
  \bibfield  {author} {\bibinfo {author} {\bibfnamefont {M.~G.}\ \bibnamefont
  {Aartsen}} \emph {et~al.} (\bibinfo {collaboration} {IceCube
  Collaboration}),\ }\href {\doibase 10/gc2bq9} {\bibfield  {journal} {\bibinfo
   {journal} {A\&A}\ }\textbf {\bibinfo {volume} {607}},\ \bibinfo {pages}
  {A115} (\bibinfo {year} {2017}{\natexlab{d}})},\ \Eprint
  {http://arxiv.org/abs/1702.06131} {arXiv:1702.06131} \BibitemShut {NoStop}%
\bibitem [{\citenamefont {Albert}\ \emph {et~al.}(2017)\citenamefont {Albert}
  \emph {et~al.}}]{antares_collaboration_search_2017}%
  \BibitemOpen
  \bibfield  {author} {\bibinfo {author} {\bibfnamefont {A.}~\bibnamefont
  {Albert}} \emph {et~al.} (\bibinfo {collaboration} {{ANTARES Collaboration}
  \& {IceCube Collaboration} \& LIGO Scientific Collaboration \& Virgo
  Collaboration}),\ }\href {\doibase 10/gbvtvn} {\bibfield  {journal} {\bibinfo
   {journal} {Phys. Rev. D}\ }\textbf {\bibinfo {volume} {96}},\ \bibinfo
  {pages} {022005} (\bibinfo {year} {2017})},\ \Eprint
  {http://arxiv.org/abs/1703.06298} {arXiv:1703.06298} \BibitemShut {NoStop}%
\bibitem [{\citenamefont {Aartsen}\ \emph
  {et~al.}(2017{\natexlab{e}})\citenamefont {Aartsen} \emph
  {et~al.}}]{aartsen_search_2017}%
  \BibitemOpen
  \bibfield  {author} {\bibinfo {author} {\bibfnamefont {M.~G.}\ \bibnamefont
  {Aartsen}} \emph {et~al.} (\bibinfo {collaboration} {IceCube
  Collaboration}),\ }\href {\doibase 10/gcpnxn} {\bibfield  {journal} {\bibinfo
   {journal} {{Astrophys. J.}}\ }\textbf {\bibinfo {volume} {846}},\ \bibinfo
  {pages} {136} (\bibinfo {year} {2017}{\natexlab{e}})},\ \Eprint
  {http://arxiv.org/abs/1705.02383} {arXiv:1705.02383} \BibitemShut {NoStop}%
\bibitem [{\citenamefont {Aartsen}\ \emph
  {et~al.}(2017{\natexlab{f}})\citenamefont {Aartsen} \emph
  {et~al.}}]{aartsen_constraints_2017}%
  \BibitemOpen
  \bibfield  {author} {\bibinfo {author} {\bibfnamefont {M.~G.}\ \bibnamefont
  {Aartsen}} \emph {et~al.} (\bibinfo {collaboration} {IceCube
  Collaboration}),\ }\href {\doibase 10/gcps5t} {\bibfield  {journal} {\bibinfo
   {journal} {Astrophys. J.}\ }\textbf {\bibinfo {volume} {849}},\ \bibinfo
  {pages} {67} (\bibinfo {year} {2017}{\natexlab{f}})},\ \Eprint
  {http://arxiv.org/abs/1707.03416} {arXiv:1707.03416} \BibitemShut {NoStop}%
\bibitem [{\citenamefont {Abbasi}\ \emph {et~al.}(2011)\citenamefont {Abbasi}
  \emph {et~al.}}]{abbasi_time-integrated_2011}%
  \BibitemOpen
  \bibfield  {author} {\bibinfo {author} {\bibfnamefont {R.}~\bibnamefont
  {Abbasi}} \emph {et~al.} (\bibinfo {collaboration} {IceCube Collaboration}),\
  }\href {\doibase 10.1088/0004-637X/732/1/18} {\bibfield  {journal} {\bibinfo
  {journal} {Astrophys. J.}\ }\textbf {\bibinfo {volume} {732}},\ \bibinfo
  {pages} {18} (\bibinfo {year} {2011})},\ \Eprint
  {http://arxiv.org/abs/1012.2137} {arXiv:1012.2137} \BibitemShut {NoStop}%
\bibitem [{\citenamefont {Aartsen}\ \emph
  {et~al.}(2013{\natexlab{b}})\citenamefont {Aartsen} \emph
  {et~al.}}]{icecube_collaboration_search_2013}%
  \BibitemOpen
  \bibfield  {author} {\bibinfo {author} {\bibfnamefont {M.~G.}\ \bibnamefont
  {Aartsen}} \emph {et~al.} (\bibinfo {collaboration} {IceCube
  Collaboration}),\ }\href {\doibase 10.1088/0004-637X/779/2/132} {\bibfield
  {journal} {\bibinfo  {journal} {Astrophys. J.}\ }\textbf {\bibinfo {volume}
  {779}},\ \bibinfo {pages} {132} (\bibinfo {year} {2013}{\natexlab{b}})},\
  \Eprint {http://arxiv.org/abs/1307.6669} {arXiv:1307.6669} \BibitemShut
  {NoStop}%
\bibitem [{\citenamefont {Aartsen}\ \emph
  {et~al.}(2014{\natexlab{a}})\citenamefont {Aartsen} \emph
  {et~al.}}]{icecube_collaboration_searches_2014}%
  \BibitemOpen
  \bibfield  {author} {\bibinfo {author} {\bibfnamefont {M.~G.}\ \bibnamefont
  {Aartsen}} \emph {et~al.} (\bibinfo {collaboration} {IceCube
  Collaboration}),\ }\href {\doibase 10/f25k7x} {\bibfield  {journal} {\bibinfo
   {journal} {Astrophys. J.}\ }\textbf {\bibinfo {volume} {796}},\ \bibinfo
  {pages} {109} (\bibinfo {year} {2014}{\natexlab{a}})},\ \Eprint
  {http://arxiv.org/abs/1406.6757} {arXiv:1406.6757} \BibitemShut {NoStop}%
\bibitem [{\citenamefont {Achterberg}\ \emph {et~al.}(2006)\citenamefont
  {Achterberg} \emph {et~al.}}]{icecube_paper}%
  \BibitemOpen
  \bibfield  {author} {\bibinfo {author} {\bibfnamefont {A.}~\bibnamefont
  {Achterberg}} \emph {et~al.} (\bibinfo {collaboration} {IceCube
  Collaboration}),\ }\href {\doibase 10.1016/j.astropartphys.2006.06.007}
  {\bibfield  {journal} {\bibinfo  {journal} {Astropart. Phys.}\ }\textbf
  {\bibinfo {volume} {26}},\ \bibinfo {pages} {155 } (\bibinfo {year}
  {2006})},\ \Eprint {http://arxiv.org/abs/astro-ph/0604450}
  {arXiv:astro-ph/0604450} \BibitemShut {NoStop}%
\bibitem [{\citenamefont {Aartsen}\ \emph
  {et~al.}(2017{\natexlab{g}})\citenamefont {Aartsen} \emph
  {et~al.}}]{detector_paper}%
  \BibitemOpen
  \bibfield  {author} {\bibinfo {author} {\bibfnamefont {M.~G.}\ \bibnamefont
  {Aartsen}} \emph {et~al.} (\bibinfo {collaboration} {IceCube
  Collaboration}),\ }\href {\doibase 10.1088/1748-0221/12/03/P03012} {\bibfield
   {journal} {\bibinfo  {journal} {J. Inst.}\ }\textbf {\bibinfo {volume}
  {12}},\ \bibinfo {pages} {P03012} (\bibinfo {year} {2017}{\natexlab{g}})},\
  \Eprint {http://arxiv.org/abs/1612.05093} {arXiv:1612.05093} \BibitemShut
  {NoStop}%
\bibitem [{\citenamefont {Abbasi}\ \emph {et~al.}(2010)\citenamefont {Abbasi}
  \emph {et~al.}}]{the_icecube_collaboration_calibration_2010}%
  \BibitemOpen
  \bibfield  {author} {\bibinfo {author} {\bibfnamefont {R.}~\bibnamefont
  {Abbasi}} \emph {et~al.} (\bibinfo {collaboration} {{IceCube}
  Collaboration}),\ }\href {\doibase 10/bnqwzn} {\bibfield  {journal} {\bibinfo
   {journal} {Nucl. Instrum. Meth. A}\ }\textbf {\bibinfo {volume} {618}},\
  \bibinfo {pages} {139} (\bibinfo {year} {2010})},\ \Eprint
  {http://arxiv.org/abs/1002.2442} {arXiv:1002.2442} \BibitemShut {NoStop}%
\bibitem [{\citenamefont {Abbasi}\ \emph {et~al.}(2009)\citenamefont {Abbasi}
  \emph {et~al.}}]{DOM_MainBoard_paper}%
  \BibitemOpen
  \bibfield  {author} {\bibinfo {author} {\bibfnamefont {R.}~\bibnamefont
  {Abbasi}} \emph {et~al.} (\bibinfo {collaboration} {{IceCube}
  Collaboration}),\ }\href {\doibase 10/cczpg4} {\bibfield  {journal} {\bibinfo
   {journal} {Nucl. Instrum. Meth. A}\ }\textbf {\bibinfo {volume} {601}},\
  \bibinfo {pages} {294} (\bibinfo {year} {2009})},\ \Eprint
  {http://arxiv.org/abs/0810.4930} {arXiv:0810.4930} \BibitemShut {NoStop}%
\bibitem [{\citenamefont {Aartsen}\ \emph
  {et~al.}(2013{\natexlab{c}})\citenamefont {Aartsen} \emph
  {et~al.}}]{ice_paper}%
  \BibitemOpen
  \bibfield  {author} {\bibinfo {author} {\bibfnamefont {M.~G.}\ \bibnamefont
  {Aartsen}} \emph {et~al.} (\bibinfo {collaboration} {{IceCube}
  Collaboration}),\ }\href {\doibase 10/f22g4d} {\bibfield  {journal} {\bibinfo
   {journal} {Nucl. Instrum. Meth. A}\ }\textbf {\bibinfo {volume} {711}},\
  \bibinfo {pages} {73} (\bibinfo {year} {2013}{\natexlab{c}})},\ \Eprint
  {http://arxiv.org/abs/1301.5361} {arXiv:1301.5361} \BibitemShut {NoStop}%
\bibitem [{\citenamefont {Ahrens}\ \emph {et~al.}(2004)\citenamefont {Ahrens}
  \emph {et~al.}}]{reco_paper}%
  \BibitemOpen
  \bibfield  {author} {\bibinfo {author} {\bibfnamefont {J.}~\bibnamefont
  {Ahrens}} \emph {et~al.} (\bibinfo {collaboration} {AMANDA Collaboration}),\
  }\href {\doibase 10.1016/j.nima.2004.01.065} {\bibfield  {journal} {\bibinfo
  {journal} {Nucl. Instrum. Meth. A}\ }\textbf {\bibinfo {volume} {524}},\
  \bibinfo {pages} {169 } (\bibinfo {year} {2004})},\ \Eprint
  {http://arxiv.org/abs/astro-ph/0407044} {arXiv:astro-ph/0407044} \BibitemShut
  {NoStop}%
\bibitem [{\citenamefont {Schatto}(2014)}]{Schatto14}%
  \BibitemOpen
  \bibfield  {author} {\bibinfo {author} {\bibfnamefont {K.}~\bibnamefont
  {Schatto}},\ }\emph {\bibinfo {title} {Stacked searches for high-energy
  neutrinos from blazars with IceCube}},\ \href
  {http://ubm.opus.hbz-nrw.de/volltexte/2014/3869/} {Ph.D. thesis},\ \bibinfo
  {school} {Johannes Gutenberg-Universit\"at} (\bibinfo {year}
  {2014})\BibitemShut {NoStop}%
\bibitem [{\citenamefont {Braun}\ \emph {et~al.}(2008)\citenamefont {Braun}
  \emph {et~al.}}]{ps_llh_paper}%
  \BibitemOpen
  \bibfield  {author} {\bibinfo {author} {\bibfnamefont {J.}~\bibnamefont
  {Braun}} \emph {et~al.},\ }\href {\doibase
  10.1016/j.astropartphys.2008.02.007} {\bibfield  {journal} {\bibinfo
  {journal} {Astropart. Phys.}\ }\textbf {\bibinfo {volume} {29}},\ \bibinfo
  {pages} {299 } (\bibinfo {year} {2008})},\ \Eprint
  {http://arxiv.org/abs/0801.1604} {arXiv:0801.1604} \BibitemShut {NoStop}%
\bibitem [{\citenamefont {Abbasi}\ \emph {et~al.}(2013)\citenamefont {Abbasi}
  \emph {et~al.}}]{abbasi_improved_2013}%
  \BibitemOpen
  \bibfield  {author} {\bibinfo {author} {\bibfnamefont {R.}~\bibnamefont
  {Abbasi}} \emph {et~al.} (\bibinfo {collaboration} {IceCube Collaboration}),\
  }\href {\doibase 10/f23s5q} {\bibfield  {journal} {\bibinfo  {journal} {Nucl.
  Instrum. Meth. A}\ }\textbf {\bibinfo {volume} {703}},\ \bibinfo {pages}
  {190} (\bibinfo {year} {2013})},\ \Eprint {http://arxiv.org/abs/1208.3430}
  {arXiv:1208.3430} \BibitemShut {NoStop}%
\bibitem [{\citenamefont {Neunh\"{o}ffer}(2006)}]{neunhoffer_estimating_2006}%
  \BibitemOpen
  \bibfield  {author} {\bibinfo {author} {\bibfnamefont {T.}~\bibnamefont
  {Neunh\"{o}ffer}},\ }\href {\doibase 10/dghq58} {\bibfield  {journal}
  {\bibinfo  {journal} {Astropart. Phys.}\ }\textbf {\bibinfo {volume} {25}},\
  \bibinfo {pages} {220} (\bibinfo {year} {2006})},\ \Eprint
  {http://arxiv.org/abs/astro-ph/0403367} {arXiv:astro-ph/0403367} \BibitemShut
  {NoStop}%
\bibitem [{\citenamefont {G\'{o}rski}\ \emph {et~al.}(2005)\citenamefont
  {G\'{o}rski} \emph {et~al.}}]{gorski_healpix_2005}%
  \BibitemOpen
  \bibfield  {author} {\bibinfo {author} {\bibfnamefont {K.~M.}\ \bibnamefont
  {G\'{o}rski}} \emph {et~al.},\ }\href {\doibase 10/bw26c9} {\bibfield
  {journal} {\bibinfo  {journal} {Astrophys. J.}\ }\textbf {\bibinfo {volume}
  {622}},\ \bibinfo {pages} {759} (\bibinfo {year} {2005})},\ \Eprint
  {http://arxiv.org/abs/astro-ph/0409513} {arXiv:astro-ph/0409513} \BibitemShut
  {NoStop}%
\bibitem [{\citenamefont {Kolmogorov}(1933)}]{kolmogorov_sulla_1933}%
  \BibitemOpen
  \bibfield  {author} {\bibinfo {author} {\bibfnamefont {A.}~\bibnamefont
  {Kolmogorov}},\ }\href@noop {} {\bibfield  {journal} {\bibinfo  {journal}
  {Inst. Ital. Attuari, Giorn.}\ }\textbf {\bibinfo {volume} {4}},\ \bibinfo
  {pages} {83} (\bibinfo {year} {1933})}\BibitemShut {NoStop}%
\bibitem [{\citenamefont {Smirnov}(1939)}]{smirnov_estimation_1939}%
  \BibitemOpen
  \bibfield  {author} {\bibinfo {author} {\bibfnamefont {N.~V.}\ \bibnamefont
  {Smirnov}},\ }\href@noop {} {\bibfield  {journal} {\bibinfo  {journal} {Bull.
  Math. Univ. Moscou}\ }\textbf {\bibinfo {volume} {2}},\ \bibinfo {pages} {3}
  (\bibinfo {year} {1939})}\BibitemShut {NoStop}%
\bibitem [{\citenamefont {Aartsen}\ \emph
  {et~al.}(2014{\natexlab{b}})\citenamefont {Aartsen} \emph
  {et~al.}}]{aartsen_energy_2014}%
  \BibitemOpen
  \bibfield  {author} {\bibinfo {author} {\bibfnamefont {M.~G.}\ \bibnamefont
  {Aartsen}} \emph {et~al.} (\bibinfo {collaboration} {{IceCube}
  Collaboration}),\ }\href {\doibase 10/f22fcd} {\bibfield  {journal} {\bibinfo
   {journal} {J. Inst.}\ }\textbf {\bibinfo {volume} {9}},\ \bibinfo {pages}
  {P03009} (\bibinfo {year} {2014}{\natexlab{b}})},\ \Eprint
  {http://arxiv.org/abs/1311.4767} {arXiv:1311.4767} \BibitemShut {NoStop}%
\bibitem [{\citenamefont {Bezrukov}\ and\ \citenamefont
  {Bugaev}(1981)}]{bezrukov_nucleon_1981}%
  \BibitemOpen
  \bibfield  {author} {\bibinfo {author} {\bibfnamefont {L.~B.}\ \bibnamefont
  {Bezrukov}}\ and\ \bibinfo {author} {\bibfnamefont {E.~V.}\ \bibnamefont
  {Bugaev}},\ }\href {https://www.osti.gov/biblio/5909968} {\bibfield
  {journal} {\bibinfo  {journal} {Sov. J. Nucl. Phys.}\ }\textbf {\bibinfo
  {volume} {33}},\ \bibinfo {pages} {1195} (\bibinfo {year}
  {1981})}\BibitemShut {NoStop}%
\bibitem [{\citenamefont {Bugaev}\ and\ \citenamefont
  {Shlepin}(2003{\natexlab{a}})}]{bugaev_photonuclear_20032}%
  \BibitemOpen
  \bibfield  {author} {\bibinfo {author} {\bibfnamefont {E.~V.}\ \bibnamefont
  {Bugaev}}\ and\ \bibinfo {author} {\bibfnamefont {Y.~V.}\ \bibnamefont
  {Shlepin}},\ }\href {\doibase 10/fhvsc9} {\bibfield  {journal} {\bibinfo
  {journal} {Nucl. Phys. B}\ }\textbf {\bibinfo {volume} {122}},\ \bibinfo
  {pages} {341} (\bibinfo {year} {2003}{\natexlab{a}})}\BibitemShut {NoStop}%
\bibitem [{\citenamefont {Bugaev}\ and\ \citenamefont
  {Shlepin}(2003{\natexlab{b}})}]{bugaev_photonuclear_2003}%
  \BibitemOpen
  \bibfield  {author} {\bibinfo {author} {\bibfnamefont {E.~V.}\ \bibnamefont
  {Bugaev}}\ and\ \bibinfo {author} {\bibfnamefont {Y.~V.}\ \bibnamefont
  {Shlepin}},\ }\href {\doibase 10/c7hhd8} {\bibfield  {journal} {\bibinfo
  {journal} {Phys. Rev. D}\ }\textbf {\bibinfo {volume} {67}},\ \bibinfo
  {pages} {034027} (\bibinfo {year} {2003}{\natexlab{b}})},\ \Eprint
  {http://arxiv.org/abs/hep-ph/0203096} {arXiv:hep-ph/0203096} \BibitemShut
  {NoStop}%
\bibitem [{\citenamefont {Bugaev}\ \emph {et~al.}(2004)\citenamefont {Bugaev},
  \citenamefont {Montaruli}, \citenamefont {Shlepin},\ and\ \citenamefont
  {Sokalski}}]{bugaev_propagation_2004}%
  \BibitemOpen
  \bibfield  {author} {\bibinfo {author} {\bibfnamefont {E.}~\bibnamefont
  {Bugaev}}, \bibinfo {author} {\bibfnamefont {T.}~\bibnamefont {Montaruli}},
  \bibinfo {author} {\bibfnamefont {Y.}~\bibnamefont {Shlepin}}, \ and\
  \bibinfo {author} {\bibfnamefont {I.}~\bibnamefont {Sokalski}},\ }\href
  {\doibase 10/d4ndx7} {\bibfield  {journal} {\bibinfo  {journal} {Astropart.
  Phys.}\ }\textbf {\bibinfo {volume} {21}},\ \bibinfo {pages} {491} (\bibinfo
  {year} {2004})},\ \Eprint {http://arxiv.org/abs/hep-ph/0312295}
  {arXiv:hep-ph/0312295} \BibitemShut {NoStop}%
\bibitem [{\citenamefont {Abramowicz}\ \emph {et~al.}(1991)\citenamefont
  {Abramowicz}, \citenamefont {Levin}, \citenamefont {Levy},\ and\
  \citenamefont {Maor}}]{abramowicz_parametrization_1991}%
  \BibitemOpen
  \bibfield  {author} {\bibinfo {author} {\bibfnamefont {H.}~\bibnamefont
  {Abramowicz}}, \bibinfo {author} {\bibfnamefont {E.~M.}\ \bibnamefont
  {Levin}}, \bibinfo {author} {\bibfnamefont {A.}~\bibnamefont {Levy}}, \ and\
  \bibinfo {author} {\bibfnamefont {U.}~\bibnamefont {Maor}},\ }\href {\doibase
  10/dnmmvw} {\bibfield  {journal} {\bibinfo  {journal} {Phys. Lett. B}\
  }\textbf {\bibinfo {volume} {269}},\ \bibinfo {pages} {465} (\bibinfo {year}
  {1991})}\BibitemShut {NoStop}%
\bibitem [{\citenamefont {Abramowicz}\ and\ \citenamefont
  {Levy}()}]{abramowicz_allm_1997}%
  \BibitemOpen
  \bibfield  {author} {\bibinfo {author} {\bibfnamefont {H.}~\bibnamefont
  {Abramowicz}}\ and\ \bibinfo {author} {\bibfnamefont {A.}~\bibnamefont
  {Levy}},\ }\href {http://arxiv.org/abs/hep-ph/9712415} {\bibfield  {journal}
  {\bibinfo  {journal} {DESY}\ }\textbf {\bibinfo {volume} {97-251}}},\ \Eprint
  {http://arxiv.org/abs/hep-ph/9712415} {arXiv:hep-ph/9712415} \BibitemShut
  {NoStop}%
\bibitem [{\citenamefont {Koehne}\ \emph {et~al.}(2013)\citenamefont {Koehne}
  \emph {et~al.}}]{koehne_proposal:_2013}%
  \BibitemOpen
  \bibfield  {author} {\bibinfo {author} {\bibfnamefont {J.-H.}\ \bibnamefont
  {Koehne}} \emph {et~al.},\ }\href {\doibase 10/f44559} {\bibfield  {journal}
  {\bibinfo  {journal} {Comput. Phys. Commun.}\ }\textbf {\bibinfo {volume}
  {184}},\ \bibinfo {pages} {2070 } (\bibinfo {year} {2013})}\BibitemShut
  {NoStop}%
\bibitem [{\citenamefont {Acero}\ \emph {et~al.}(2015)\citenamefont {Acero}
  \emph {et~al.}}]{3FGL}%
  \BibitemOpen
  \bibfield  {author} {\bibinfo {author} {\bibfnamefont {F.}~\bibnamefont
  {Acero}} \emph {et~al.} (\bibinfo {collaboration} {Fermi-LAT
  Collaboration}),\ }\href {\doibase 10.1088/0067-0049/218/2/23} {\bibfield
  {journal} {\bibinfo  {journal} {Astrophys. J. Suppl.}\ }\textbf {\bibinfo
  {volume} {218}},\ \bibinfo {pages} {23} (\bibinfo {year} {2015})},\ \Eprint
  {http://arxiv.org/abs/1501.02003} {arXiv:1501.02003} \BibitemShut {NoStop}%
%%CITATION = ARXIV:1501.02003;%%
\bibitem [{\citenamefont {Ajello}\ \emph {et~al.}(2017)\citenamefont {Ajello}
  \emph {et~al.}}]{3FHL}%
  \BibitemOpen
  \bibfield  {author} {\bibinfo {author} {\bibfnamefont {M.}~\bibnamefont
  {Ajello}} \emph {et~al.} (\bibinfo {collaboration} {Fermi-LAT
  Collaboration}),\ }\href {\doibase 10.3847/1538-4365/aa8221} {\bibfield
  {journal} {\bibinfo  {journal} {Astrophys. J. Suppl.}\ }\textbf {\bibinfo
  {volume} {232}},\ \bibinfo {pages} {18} (\bibinfo {year} {2017})},\ \Eprint
  {http://arxiv.org/abs/1702.00664} {arXiv:1702.00664} \BibitemShut {NoStop}%
%%CITATION = ARXIV:1702.00664;%%
\bibitem [{\citenamefont {Wilks}(1938)}]{wilks_large-sample_1938}%
  \BibitemOpen
  \bibfield  {author} {\bibinfo {author} {\bibfnamefont {S.~S.}\ \bibnamefont
  {Wilks}},\ }\href {\doibase 10/fktrn4} {\bibfield  {journal} {\bibinfo
  {journal} {Ann. Math. Statist.}\ }\textbf {\bibinfo {volume} {9}},\ \bibinfo
  {pages} {60} (\bibinfo {year} {1938})}\BibitemShut {NoStop}%
\bibitem [{\citenamefont {Glauch}\ and\ \citenamefont
  {Turcati}(2017)}]{glauchsearch2017}%
  \BibitemOpen
  \bibfield  {author} {\bibinfo {author} {\bibfnamefont {T.}~\bibnamefont
  {Glauch}}\ and\ \bibinfo {author} {\bibfnamefont {A.}~\bibnamefont {Turcati}}
  (\bibinfo {collaboration} {IceCube Collaboration}),\ }\href
  {https://pos.sissa.it/301/1014/} {\bibfield  {journal} {\bibinfo  {journal}
  {PoS}\ }\textbf {\bibinfo {volume} {ICRC2017}},\ \bibinfo {pages} {1014}
  (\bibinfo {year} {2017})},\ \Eprint {http://arxiv.org/abs/1710.01179}
  {arXiv:1710.01179} \BibitemShut {NoStop}%
\bibitem [{\citenamefont {Neyman}(1937)}]{neyman_upper_limit}%
  \BibitemOpen
  \bibfield  {author} {\bibinfo {author} {\bibfnamefont {J.}~\bibnamefont
  {Neyman}},\ }\href {\doibase 10.1098/rsta.1937.0005} {\bibfield  {journal}
  {\bibinfo  {journal} {Phil. Trans. R. Soc. A}\ }\textbf {\bibinfo {volume}
  {236}},\ \bibinfo {pages} {333} (\bibinfo {year} {1937})}\BibitemShut
  {NoStop}%
\bibitem [{\citenamefont {{IceCube
  Collaboration}}(2017)}]{icecube_collaboration_amon}%
  \BibitemOpen
  \bibfield  {author} {\bibinfo {author} {\bibnamefont {{IceCube
  Collaboration}}},\ }\href {https://gcn.gsfc.nasa.gov/gcn3/21916.gcn3}
  {\bibfield  {journal} {\bibinfo  {journal} {GRB Coordinates Network, Circular
  Service}\ }\textbf {\bibinfo {volume} {21916}} (\bibinfo {year}
  {2017})}\BibitemShut {NoStop}%
\bibitem [{\citenamefont {Taboada}\ \emph {et~al.}(2017)\citenamefont
  {Taboada}, \citenamefont {Tung},\ and\ \citenamefont
  {Wood}}]{taboada_constrains_2018}%
  \BibitemOpen
  \bibfield  {author} {\bibinfo {author} {\bibfnamefont {I.}~\bibnamefont
  {Taboada}}, \bibinfo {author} {\bibfnamefont {C.~F.}\ \bibnamefont {Tung}}, \
  and\ \bibinfo {author} {\bibfnamefont {J.}~\bibnamefont {Wood}} (\bibinfo
  {collaboration} {HAWC Collaboration}),\ }\href
  {https://pos.sissa.it/301/663/} {\bibfield  {journal} {\bibinfo  {journal}
  {PoS}\ }\textbf {\bibinfo {volume} {ICRC2017}},\ \bibinfo {pages} {663}
  (\bibinfo {year} {2017})},\ \Eprint {http://arxiv.org/abs/1801.09545}
  {arXiv:1801.09545} \BibitemShut {NoStop}%
\bibitem [{\citenamefont {Murase}\ and\ \citenamefont
  {Waxman}(2016)}]{murase_constraining_2016}%
  \BibitemOpen
  \bibfield  {author} {\bibinfo {author} {\bibfnamefont {K.}~\bibnamefont
  {Murase}}\ and\ \bibinfo {author} {\bibfnamefont {E.}~\bibnamefont
  {Waxman}},\ }\href {\doibase 10.1103/PhysRevD.94.103006} {\bibfield
  {journal} {\bibinfo  {journal} {Phys. Rev. D}\ }\textbf {\bibinfo {volume}
  {94}},\ \bibinfo {pages} {103006} (\bibinfo {year} {2016})},\ \Eprint
  {http://arxiv.org/abs/1607.01601} {arXiv:1607.01601} \BibitemShut {NoStop}%
\bibitem [{\citenamefont {Hopkins}\ and\ \citenamefont
  {Beacom}(2006)}]{hopkins_normalization_2006}%
  \BibitemOpen
  \bibfield  {author} {\bibinfo {author} {\bibfnamefont {A.~M.}\ \bibnamefont
  {Hopkins}}\ and\ \bibinfo {author} {\bibfnamefont {J.~F.}\ \bibnamefont
  {Beacom}},\ }\href {\doibase 10.1086/506610} {\bibfield  {journal} {\bibinfo
  {journal} {Astrophys. J.}\ }\textbf {\bibinfo {volume} {651}},\ \bibinfo
  {pages} {142} (\bibinfo {year} {2006})},\ \Eprint
  {http://arxiv.org/abs/astro-ph/0601463} {arXiv:astro-ph/0601463} \BibitemShut
  {NoStop}%
\bibitem [{\citenamefont {Ade}\ \emph {et~al.}(2016)\citenamefont {Ade} \emph
  {et~al.}}]{planck_collaboration_planck_2016}%
  \BibitemOpen
  \bibfield  {author} {\bibinfo {author} {\bibfnamefont {P.~A.~R.}\
  \bibnamefont {Ade}} \emph {et~al.} (\bibinfo {collaboration} {Planck
  Collaboration}),\ }\href {\doibase 10/f9scmm} {\bibfield  {journal} {\bibinfo
   {journal} {A\&A}\ }\textbf {\bibinfo {volume} {594}},\ \bibinfo {pages}
  {A13} (\bibinfo {year} {2016})},\ \Eprint {http://arxiv.org/abs/1502.01589}
  {arXiv:1502.01589} \BibitemShut {NoStop}%
\bibitem [{\citenamefont {Amato}\ \emph {et~al.}(2003)\citenamefont {Amato},
  \citenamefont {Guetta},\ and\ \citenamefont {Blasi}}]{amatosignatures2003}%
  \BibitemOpen
  \bibfield  {author} {\bibinfo {author} {\bibfnamefont {E.}~\bibnamefont
  {Amato}}, \bibinfo {author} {\bibfnamefont {D.}~\bibnamefont {Guetta}}, \
  and\ \bibinfo {author} {\bibfnamefont {P.}~\bibnamefont {Blasi}},\ }\href
  {\doibase 10.1051/0004-6361:20030279} {\bibfield  {journal} {\bibinfo
  {journal} {A\&A}\ }\textbf {\bibinfo {volume} {402}},\ \bibinfo {pages} {827}
  (\bibinfo {year} {2003})}\BibitemShut {NoStop}%
\bibitem [{\citenamefont {Kappes}\ \emph {et~al.}(2007)\citenamefont {Kappes},
  \citenamefont {Hinton}, \citenamefont {Stegmann},\ and\ \citenamefont
  {Aharonian}}]{kappespotential2007}%
  \BibitemOpen
  \bibfield  {author} {\bibinfo {author} {\bibfnamefont {A.}~\bibnamefont
  {Kappes}}, \bibinfo {author} {\bibfnamefont {J.}~\bibnamefont {Hinton}},
  \bibinfo {author} {\bibfnamefont {C.}~\bibnamefont {Stegmann}}, \ and\
  \bibinfo {author} {\bibfnamefont {F.~A.}\ \bibnamefont {Aharonian}},\ }\href
  {\doibase 10.1086/518161} {\bibfield  {journal} {\bibinfo  {journal}
  {Astrophys. J.}\ }\textbf {\bibinfo {volume} {661}},\ \bibinfo {pages} {1348}
  (\bibinfo {year} {2007})},\ \Eprint {http://arxiv.org/abs/astro-ph/0607286}
  {arXiv:astro-ph/0607286} \BibitemShut {NoStop}%
\bibitem [{\citenamefont {Reimer}(2015)}]{reimerphotonneutrino}%
  \BibitemOpen
  \bibfield  {author} {\bibinfo {author} {\bibfnamefont {A.}~\bibnamefont
  {Reimer}},\ }\href {\doibase 10.22323/1.236.1123} {\bibfield  {journal}
  {\bibinfo  {journal} {PoS}\ }\textbf {\bibinfo {volume} {ICRC2015}},\
  \bibinfo {pages} {1123} (\bibinfo {year} {2015})}\BibitemShut {NoStop}%
\bibitem [{\citenamefont {Petropoulou}\ \emph {et~al.}(2015)\citenamefont
  {Petropoulou}, \citenamefont {Dimitrakoudis}, \citenamefont {Padovani},
  \citenamefont {Mastichiadis},\ and\ \citenamefont
  {Resconi}}]{petropoulouphotohadronic2015}%
  \BibitemOpen
  \bibfield  {author} {\bibinfo {author} {\bibfnamefont {M.}~\bibnamefont
  {Petropoulou}}, \bibinfo {author} {\bibfnamefont {S.}~\bibnamefont
  {Dimitrakoudis}}, \bibinfo {author} {\bibfnamefont {P.}~\bibnamefont
  {Padovani}}, \bibinfo {author} {\bibfnamefont {A.}~\bibnamefont
  {Mastichiadis}}, \ and\ \bibinfo {author} {\bibfnamefont {E.}~\bibnamefont
  {Resconi}},\ }\href {\doibase 10.1093/mnras/stv179} {\bibfield  {journal}
  {\bibinfo  {journal} {Mon. Not. R. Astron. Soc.}\ }\textbf {\bibinfo {volume}
  {448}},\ \bibinfo {pages} {2412} (\bibinfo {year} {2015})},\ \Eprint
  {http://arxiv.org/abs/1501.07115} {arXiv:1501.07115} \BibitemShut {NoStop}%
\bibitem [{\citenamefont {Mandelartz}\ and\ \citenamefont
  {Becker~Tjus}(2015)}]{mandelartzprediction2015}%
  \BibitemOpen
  \bibfield  {author} {\bibinfo {author} {\bibfnamefont {M.}~\bibnamefont
  {Mandelartz}}\ and\ \bibinfo {author} {\bibfnamefont {J.}~\bibnamefont
  {Becker~Tjus}},\ }\href {\doibase 10.1016/j.astropartphys.2014.12.002}
  {\bibfield  {journal} {\bibinfo  {journal} {Astropart. Phys.}\ }\textbf
  {\bibinfo {volume} {65}},\ \bibinfo {pages} {80} (\bibinfo {year} {2015})},\
  \Eprint {http://arxiv.org/abs/1301.2437} {arXiv:1301.2437} \BibitemShut
  {NoStop}%
\bibitem [{\citenamefont {Aleksi\'c}\ \emph
  {et~al.}(2015{\natexlab{a}})\citenamefont {Aleksi\'c} \emph
  {et~al.}}]{aleksic_measurement_2015}%
  \BibitemOpen
  \bibfield {author} {\bibinfo {author} {\bibfnamefont {J.}\ \bibnamefont
  {Aleksi\'c}} \emph {et~al.} (\bibinfo {collaboration} {MAGIC
  Collaboration}),\ }\href {\doibase 10/gd8vwt} {\bibfield
  {journal} {\bibinfo  {journal} {JHEAp}\ }\textbf {\bibinfo {volume}
  {5-6}},\ \bibinfo {pages} {30} (\bibinfo {year} {2015}{\natexlab{a}})},\
  \Eprint {http://arxiv.org/abs/1406.6892} {arXiv:1406.6892} \BibitemShut
  {NoStop}%
\bibitem [{\citenamefont {Aharonian}\ \emph {et~al.}(2006)\citenamefont
  {Aharonian} \emph {et~al.}}]{aharonian_observations_2006}%
  \BibitemOpen
  \bibfield  {author} {\bibinfo {author} {\bibfnamefont {F.}~\bibnamefont
  {Aharonian}} \emph {et~al.} (\bibinfo {collaboration} {HESS Collaboration}),\
  }\href {\doibase 10/c3j279} {\bibfield  {journal} {\bibinfo  {journal}
  {A\&A}\ }\textbf {\bibinfo {volume} {457}},\ \bibinfo {pages} {899} (\bibinfo
  {year} {2006})},\ \Eprint {http://arxiv.org/abs/astro-ph/0607333}
  {arXiv:astro-ph/0607333} \BibitemShut {NoStop}%
\bibitem [{\citenamefont {Aliu}\ \emph {et~al.}(2006)\citenamefont
  {Aliu} \emph {et~al.}}]{aliu_investigating_2014}%
  \BibitemOpen 
  \bibfield {author} {\bibinfo {author} {\bibfnamefont {E.}~\bibnamefont
  {Aliu}} \emph {et~al.} (\bibinfo {collaboration} {VERITAS Collaboration}),\
  }\href {\doibase 10/f3n8mj} {\bibfield  {journal} {\bibinfo  {journal}
  {Astrophys. J.}\ }\textbf {\bibinfo {volume} {787}},\ \bibinfo {pages} {166} (\bibinfo
  {year} {2014})},\ \Eprint {http://arxiv.org/abs/1404.7185}
  {arXiv:1404.7185} \BibitemShut {NoStop}%
\end{thebibliography}

\appendix

\section{Performance of individual sub-samples}

\begin{figure}
    \centering
    \input{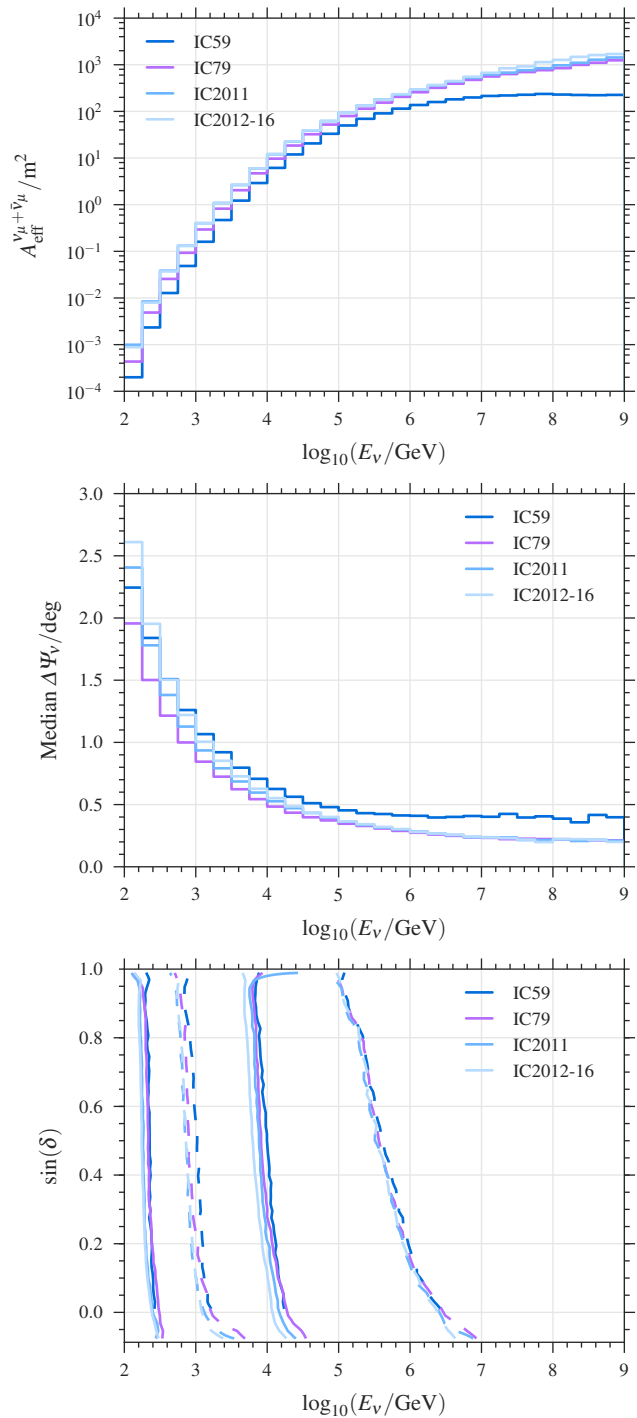}
    \caption{Top: Muon neutrino and anti-neutrino effective area averaged over the Northern hemisphere as function of $\log_{10}$ of neutrino energy. Middle: Median neutrino angular resolution as function of $\log_{10}$ of neutrino energy. Bottom: Central 90\% neutrino energy range for atmospheric (astrophysical) neutrinos as solid (dashed) line for each declination. 
    Lines are labeled by there sub-season.
    \label{fig:effective_area_seasons} \label{fig:angular_resolution_seasons} \label{fig:erange_seasons}}
\end{figure}

The quality and statistical power of a sample, w.r.t. a search for point-like sources, can be characterized by the effective area of muon-neutrino and anti-neutrino detection, the point spread function and the central 90\% energy range (see Section~\ref{sec:datasample}). As the data were taken with different partial configurations of IceCube, the details of the event selections are different for each season. In Fig.~\ref{fig:effective_area} the livetime average of all sub-samples is shown. In Fig.~\ref{fig:effective_area_seasons} the effective area, point spread function and central 90\% energy range are shown for each sub-sample individually. The plot shows that - despite of different detector configurations and event selections - the characteristics of the event samples are similar.

\section{Results for diffuse best-fit spectral index}

\begin{figure}[!htb]
        \centering
        \input{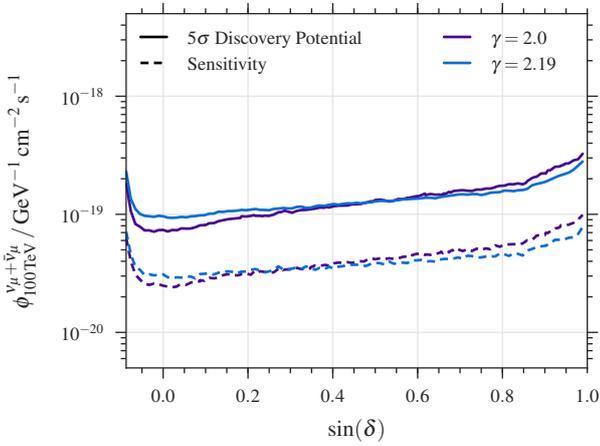}
        \caption{Sensitivity and discovery potential on the flux normalization at $100\,\mathrm{TeV}$ for an $E^{-\gamma}$ power-law spectrum.
                 Lines are given for $\gamma=2.0$ as in Fig.~\ref{fig:sens_bandwidth} and the diffuse best-fit spectral index of $\gamma=2.19$.}
        \label{fig:sensitivity_different_gamma}
\end{figure}

An $E^{-2}$ power-law is often used as benchmark model and for a comparison between publications.
However, the diffuse best-fit spectral index is $\gamma=2.19$, which is, given the uncertainties is not consistent with $\gamma=2$. Therefore, the sensitivity and discovery potential
for single point sources are recalculated using this spectral index. In Fig.~\ref{fig:sensitivity_different_gamma},
the sensitivity and discovery potential for an $E^{-\gamma}$ spectum are shown with $\gamma=2.0$ and $\gamma=2.19$. 
The flux normalization is evaluated at a pivot energy of $100\,\mathrm{TeV}$. 
The sensitivity and discovery potential for the assumed spectral indices turn out to be very similar. 

\begin{figure}[!htb]
        \centering
        \input{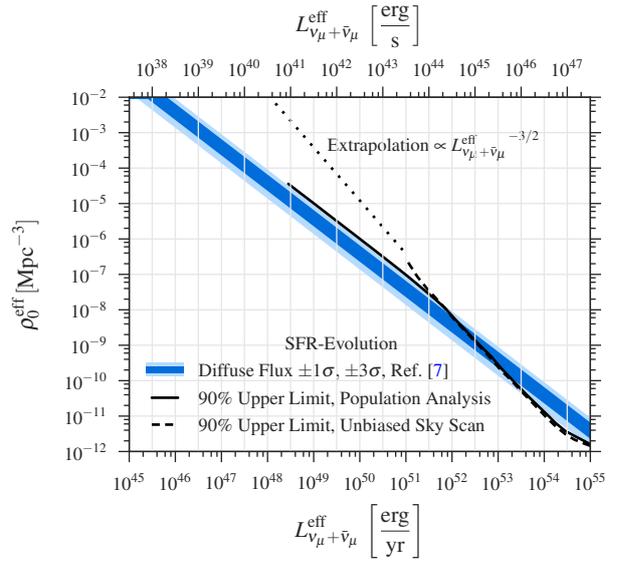}
        \caption{Same as Fig.~\ref{fig:HPA_UL_Kowalski} but for $\gamma=2.19$.}
        \label{fig:kowalski_2_19}
\end{figure}

In addition also the 90\% CL upper limit on source populations as described in Section~\ref{sec:Kowalski} are recalculated for a spectral index of $\gamma=2.19$.
The upper limit are shown in Fig.~\ref{fig:kowalski_2_19}. Comparing with Fig.~\ref{fig:HPA_UL_Kowalski}, there is no strong indication of a dependence on the spectral index.

\end{document}